\documentclass[twocolumn,dvipsnames]{aastex62}
\pdfoutput=1 
\usepackage{amsmath,amstext}
\usepackage{apjfonts}
\usepackage[figure,figure*]{hypcap}
\usepackage{ulem}
\usepackage{xspace}

\usepackage{xfrac}
\usepackage{graphicx}
\usepackage{natbib}
\usepackage{amssymb}
\usepackage{amsmath}
\usepackage{array,booktabs}
\usepackage{mathptmx}
\usepackage{booktabs}
\usepackage{hyperref}
\usepackage{float}
\usepackage{todonotes}
\newcommand{\mej}{\,{M_{\rm ej}}}
\newcommand{\msun}{\,{M_{\odot}}}
\newcommand{\mbh}{\,{M_{\rm BH}}}

\DeclareSymbolFont{cmletters}{OML}{cmm}{m}{it}
\DeclareMathSymbol{v}{\mathalpha}{cmletters}{"76}

\shorttitle{Relativistic Jets from BH--NS Mergers}
\shortauthors{Gottlieb et al.}

\begin{document}
\title{Large-scale Evolution of Seconds-long Relativistic Jets from Black Hole--Neutron Star Mergers}

    \author[0000-0003-3115-2456]{Ore Gottlieb$^*$}
	\email{oregottlieb@gmail.com}
	\affiliation{Center for Interdisciplinary Exploration \& Research in Astrophysics (CIERA), Physics \& Astronomy, Northwestern University, Evanston, IL 60202, USA}

    \author[0009-0005-2478-7631]{Danat Issa$^*$}
	\affiliation{Center for Interdisciplinary Exploration \& Research in Astrophysics (CIERA), Physics \& Astronomy, Northwestern University, Evanston, IL 60202, USA}

    \author[0000-0003-2982-0005]{Jonatan Jacquemin-Ide}
	\affiliation{Center for Interdisciplinary Exploration \& Research in Astrophysics (CIERA), Physics \& Astronomy, Northwestern University, Evanston, IL 60202, USA}
 
    \author[0000-0003-4475-9345]{Matthew Liska}
    \affiliation{Institute for Theory and Computation, Harvard University, 60 Garden Street, Cambridge, MA 02138, USA; John Harvard Distinguished Science and ITC}

     \author[0000-0003-4617-4738]{Francois Foucart}
	\affiliation{Department of Physics and Astronomy, University of New Hampshire, 9 Library Way, Durham, NH 03824, USA}

    \author[0000-0002-9182-2047]{Alexander Tchekhovskoy}
	\affiliation{Center for Interdisciplinary Exploration \& Research in Astrophysics (CIERA), Physics \& Astronomy, Northwestern University, Evanston, IL 60202, USA}

    \author[0000-0002-3635-5677]{Brian D. Metzger}
    \affiliation{Department of Physics and Columbia Astrophysics Laboratory, Columbia University, Pupin Hall, New York, NY 10027, USA}
    \affiliation{Center for Computational Astrophysics, Flatiron Institute, New York, NY 10010, USA}

    \author[0000-0001-9185-5044]{Eliot Quataert}
    \affiliation{Department of Astrophysical Sciences, Princeton University, Princeton, NJ 08544, USA}

    \author[0000-0002-3635-5677]{Rosalba Perna}
    \affiliation{Department of Physics and Astronomy, Stony Brook University, Stony Brook, NY 11794-3800, USA}
    \affiliation{Center for Computational Astrophysics, Flatiron Institute, New York, NY 10010, USA}

    \author[0000-0002-5981-1022]{Daniel Kasen}
    \affiliation{Astronomy Department and Theoretical Astrophysics Center, University of California, Berkeley, Berkeley, CA 94720, USA}
    \affiliation{Physics Department, University of California, Berkeley, Berkeley, CA 94720, USA}
    \affiliation{Nuclear Science Division, Lawrence Berkeley National Laboratory, Berkeley, CA 94720, USA}

    \author[0000-0002-0050-1783]{Matthew D. Duez}
    \affiliation{Department of Physics \& Astronomy, Washington State University, Pullman, Washington 99164, USA}

    \author[0000-0001-5392-7342]{Lawrence E. Kidder}
    \affiliation{Cornell Center for Astrophysics and Planetary Science, Cornell University, Ithaca, New York, 14853, USA}

    \author[0000-0001-9288-519X]{Harald P. Pfeiffer}
    \affiliation{Max Planck Institute for Gravitational Physics (Albert Einstein Institute), D-14467 Potsdam, Germany}

    \author[0000-0001-6656-9134]{Mark A. Scheel}
    \affiliation{TAPIR, Walter Burke Institute for Theoretical Physics, MC 350-17, California Institute of Technology, Pasadena, California 91125, USA}

\def\thefootnote{*}\footnotetext{These authors contributed equally to this work.}

\begin{abstract}

We present the first numerical simulations that track the evolution of a black hole--neutron star (BH--NS) merger from pre-merger to $r\gtrsim10^{11}\,{\rm cm}$. The disk that forms after a merger of mass ratio $q=2$ ejects massive disk winds ($3--5\times10^{-2}\,\msun$). We introduce various post-merger magnetic configurations and find that initial poloidal fields lead to jet launching shortly after the merger. The jet maintains a constant power due to the constancy of the large-scale BH magnetic flux until the disk becomes magnetically arrested (MAD), where the jet power falls off as $L_j\sim t^{-2}$. All jets inevitably exhibit either excessive luminosity due to rapid MAD activation when the accretion rate is high or excessive duration due to delayed MAD activation compared to typical short gamma-ray bursts (sGRBs). This provides a natural explanation for long sGRBs such as GRB 211211A but also raises a fundamental challenge to our understanding of jet formation in binary mergers. One possible implication is the necessity of higher binary mass ratios or moderate BH spins to launch typical sGRB jets. For post-merger disks with a toroidal magnetic field, dynamo processes delay jet launching such that the jets break out of the disk winds after several seconds. We show for the first time that sGRB jets with initial magnetization $\sigma_0>100$ retain significant magnetization ($\sigma\gg1$) at $r>10^{10}\,{\rm cm}$, emphasizing the importance of magnetic processes in the prompt emission. The jet-wind interaction leads to a power-law angular energy distribution by inflating an energetic cocoon whose emission is studied in a companion paper.

\end{abstract}
	
\section{Introduction}\label{sec:introduction}

The most recent observing run of LIGO--Virgo--KAGRA (LVK), O3b, yielded the detection of at least one gravitational wave (GW) source originating from a black hole--neutron star (BH--NS) merger: GW200115, with GW200105 being another controversial astrophysical source \citep{Abbott2021}. These events exhibited similar characteristics, including inferred mass ratios of $4 \lesssim q \lesssim 5$ and BH spins consistent with zero, although the BH spin in GW200115 is not well constrained. While no electromagnetic counterparts were detected for these BH--NS mergers \citep{Dichiara2021,Zhu2021}, it is still uncertain whether they are representative of the broader population of BH--NS mergers. The ongoing LVK run O4 holds the potential for the first detection of a multi-messenger BH--NS merger \citep{Abbott2020}. Similar to the case of the binary neutron star (BNS) merger GW170817 \citep[see][for reviews]{Nakar2019,Margutti2021}, BH--NS mergers can give rise to two types of electromagnetic counterparts: kilonovae and jet--cocoon emission \citep[e.g.,][]{Paczynski1991,Mochkovitch1993,Janka1999,Rosswog2005,Surman2008,Metzger2010,Tanaka2014,Fernandez2015,Fernandez2017,Foucart2015,Kawaguchi2016,Darbha2021,Wanajo2022,Ekanger2023,Gompertz2023}.

The ability to generate relativistic jets in BH--NS mergers is contingent upon various characteristics of the system, including the mass ratio, BH spin, NS radius, and spin-orbit misalignment of the binary. When the pre-merger BH spins rapidly, the misalignment is modest, the NS is not overly compact, and the mass ratio is not excessively high, a substantial amount of mass remains outside the BH innermost stable circular orbit (ISCO), facilitating the formation of a massive accretion disk around it \citep{Shibata2006,Shibata2007,Etienne2008,Rantsiou2008,Shibata2008,Shibata2011,Duez2010,Foucart2011,Foucart2012b,Foucart2014,Foucart2017,Foucart2019,Kyutoku2011,Kyutoku2013,Kyutoku2015,Foucart2012a,Kawaguchi2015,Fragione2021,Hayashi2021,Sarin2022,Biscoveanu2023}. As a result, in some cases mass continues to accrete onto the BH following the merger. If the disk brings vertical large-scale magnetic flux to the post-merger BH, the Blandford-Znajek effect leads to formation of a pair of ultra-relativistic collimated outflows, or jets \citep{Blandford1977}.

The properties of the launched jets are influenced by several factors, including the magnetic flux threading the BH, the spin of the post-merger BH, and the mass accretion rate \citep{Blandford1977,Tchekhovskoy2011,2015ASSL..414...45T}. While the BH spin can be inferred from GW observations, the mass accretion rate and the magnetic field structure cannot, motivating the need for numerical simulations to explore these properties. Both the mass accretion rate and the magnetic field threading the BH strongly depend on the disk magnetic configuration \citep{Rosswog2007}, which, in turn, relies on the amplification of the magnetic field before, during, and shortly after the merger. Various mechanisms have been proposed to generate a strong magnetic field in the context of BNS mergers. One such mechanism is the Kelvin-Helmholtz instability \citep[KHI;][]{Helmholtz1868,Thomson1871} in the shear layer between the NSs \citep{Price2006}. However, in BH--NS mergers, KHI can only occur in the shear between the spiral arm and the disk \citep{Hayashi2022a}. Another mechanism involves the magnetorotational instability (MRI) and magnetic winding in the disk. In this case, the amplification of magnetic fields in the disk through the MRI should persist until the fields reach equipartition values, where the typical thermal pressure in the disk corresponds to a magnetic field strength of $\sim 10^{15}$ G \citep{Kiuchi2015}. We note that both of these amplification processes are highly sensitive to the grid resolution and may not be fully captured in present-day simulations \citep{Kiuchi2018}.

Irrespective of the specific jet properties, it is expected that at least some jets ultimately generate a short gamma-ray burst (sGRB), followed by a multi-band afterglow emission. In a significant fraction of sGRBs \citep[$ \sim 25\% - 50\% $;][]{Norris2008,Norris2010}, there is evidence of a distinct third X-ray component known as the extended emission \citep[EE;][]{Norris2006,Perley2009}, the origin of which is still a subject of debate. One intriguing possibility is that sGRB-EEs originate from late-time fallback accretion of the tidal tail formed by the merger ejecta \citep{Rosswog2007}, if not suppressed by disk winds \citep{Fernandez2015,Fernandez2017}.

In the past decade, significant progress has been made in our understanding of the formation of relativistic jets in BH--NS mergers through numerical simulations \citep{Etienne2012,Kiuchi2015,Paschalidis2015,Ruiz2018,Hayashi2022a,Hayashi2023}. However, previous approaches were limited in their ability to track the outflows up to the self-similar expansion radii and investigate the late-time evolution of the disk and relativistic outflows. Furthermore, the structure of the relativistic outflows, which has been shown to be crucial in shaping the afterglow emission in GW170817 \citep[e.g.,][]{Alexander2017,Margutti2017,Mooley2018a,Mooley2018b,Lazzati2018,Lyman2018,Troja2018,Ghirlanda2019,Lamb2019}, remains unexplored in the context of BH--NS mergers. In this study, we present the first simulations of BH--NS mergers that extend out to radii of $r > 10^{11}\,{\rm cm} $ by chaining a numerical relativity simulation of a BH--NS merger with a suite of general relativistic magnetohydrodynamic (GRMHD) simulations. By employing this approach, we investigate the launching and evolution of sub- and ultra-relativistic outflows from BH--NS mergers under various magnetic field configurations in the post-merger disk.

\section{Setup}\label{sec:setup}

We simulate, for the first time, the entire dynamical evolution of a compact object merger from the pre-merger phase to the stage where most of the gas reaches homologous expansion. We achieve this by first performing a numerical relativity simulation using the code SpEC \citep{spec} of the pre-merger until 8 ms post-merger. We remap the output of the simulation to use it as the initial conditions for the GPU-accelerated GRMHD code \textsc{h-amr} \citep{Liska2022}. We then add different configurations of magnetic field and evolve the system for several seconds.

For the merger simulation, we consider optimal binary properties for producing a massive post-merger disk. The binary mass ratio is $ q = 2 $, where the NS gravitational mass is $M_{\rm NS}=1.35M_\odot$, and the BH Christodoulou mass is $M_{\rm BH}=2.7M_\odot$. The matter inside the NS is described by the SFHo equation of state~\citep{2013ApJ...774...17S}, and the NS is initially non-spinning. The BH has an initial dimensionless spin $a=0.6$, aligned with the orbital angular momentum of the binary. We follow the late inspiral of the binary from an initial separation $d\sim 60\,{\rm km}$, the disruption of the NS, and formation of a tidal tail and accretion disk, ending the SpEC simulation $8\,{\rm ms}$ after NS disruption. At this time, a baryonic mass of $M_{\rm rem}=0.17\,{\rm M_\odot}$ remains outside of a BH of mass $M_{\rm BH}=3.80M_\odot$ and dimensionless spin $a=0.86$. About $ \sim 0.14\,\msun $ of the baryonic mass is in a nearly Keplerian disk, and $ \sim 0.03\,\msun $ is outside of the disk, with $ \sim~1--2\times 10^{-3}\,\msun $ of which is the unbound tidal tail. The remaining matter has average temperature $\langle T\rangle = 4\,{\rm MeV}$ and electron fraction $\langle Y_e\rangle =0.06$.

The SpEC simulation is performed by evolving Einstein's equations in the generalized harmonics formalism~\citep{Lindblom:2007} coupled to the relativistic hydrodynamics equations and a Monte-Carlo scheme for neutrino radiation transport \citep[see][for a description of the numerical methods in SpEC]{Duez:2008rb,Foucart:2013a,Foucart:2021mcb}. The SpEC simulations are performed with a resolution $\Delta x = 190\,{\rm m}$ on the finest finite difference grid used to evolve the fluid equations. After the merger, that finite difference grid uses adaptive mesh refinement (AMR) with six nested blocks. Each block has $252^3$ grid cells, with the grid spacing increasing by a factor of $2$ between nested blocks. Thus, the outermost block has sides of length $L\sim 1500\,{\rm km}$, sufficient to follow the ejected matter to the end of the SpEC simulation: the total mass loss at the outer boundary over the course of our simulation is less than $10^{-3}\,M_\odot$. Given the low impact of neutrino transport on the inspiral, disruption, and early post-merger evolution, our Monte-Carlo scheme uses only $10^6$ packets per neutrino species, yet each packet has energy $\lesssim 2 \times 10^{-11} M_\odot c^2$. The Monte-Carlo radiation transport here is mostly useful to follow the evolution of the post-merger disk composition in preparation for future post-merger simulations including neutrino transport.

At 8 ms after the merger, the space-time metric is approximately axisymmetric and hardly changes, allowing us to remap the output to \textsc{h-amr}, which works with a fixed space-time metric (see Appendix~\ref{sec:remap}). Upon remmaping, we modify the equation of state from tabulated to ideal gas. We consider five models with different seed magnetic field strengths and geometries, which are summarized in Table~\ref{tab:models}. The magnetic field depends on the mass density distribution at 8 ms, with a cutoff at $ 5\times 10^{-4} $ of the maximum comoving density, $ \rho $. Poloidal magnetic field models,  $P_s$, $P_c$, and $P_w$, with ``strong'', ``canonical'', and ``weak'' magnetic field strengths, respectively, are all initialized with a poloidal magnetic field configuration with varying $ \beta_p \equiv p_g/p_m $, where $ p_g $ is the thermal pressure, and $ p_m $ is the magnetic pressure, and Table~\ref{tab:models} gives typical $\beta_p$ values. The toroidal strong magnetic field model, $T_s$, is initialized with a toroidal magnetic field, and the model $H_0$ has no magnetic field. 

We now provide more information on the magnetic field structure of the models. The model $P_w$ has a characteristic initial $\beta_p = 1000$ and a large field loop set using a magnetic field potential $A_\varphi \propto \rho^2 r^3$. The models $P_c$ and $P_s$ have the same initial field geometries, set by $A_\varphi \propto \rho$, with a characteristic $\beta_p$ set to 1000 and 100, respectively. The model $T_s$ is initialized with a rather strong toroidal field with characteristic $\beta_p = 1$ and $A_\theta \propto \rho$. Figure~\ref{fig:betap_ini} in Appendix~\ref{sec:remap} gives the initial $ \beta_p $ profiles. If jets are launched, their initial magnetization, set by the density floor conditions, is $ \sigma_0 \equiv b_0^2/4\pi\rho_0 c^2 = 150 $, where $ b_0 $ is the initial comoving magnetic field strength, and $ \rho_0 $ is the initial comoving mass density. This is the first time that such high initial magnetizations are explored in the context of binary mergers.

	\begin{table}
		\setlength{\tabcolsep}{2.9pt}
		\centering
		\renewcommand{\arraystretch}{1.2}
		\begin{tabular}{| c | c c c c | c c | }
			
                \hline
			Model & $ A $ & $ \beta_p $ & $ {\rm log}(B\,[{\rm G}]) $ & $ t_f\,[{\rm s}] $ & $ \mej\,[10^{-2}\,\msun] $ & $ t_b\,[{\rm s}] $
			\\	\hline
			$ H_0 $ & $ A = 0 $ & - & - & $ 8 $ & 3 & - \\ %BHNS_B0
			$ P_w $ & $ A_\varphi \propto \rho^2r^3 $ & 1000 & 15 & $ 5 $ & 3 & 0.3\\ %BHNSh
			$ P_c $ & $ A_\varphi \propto \rho $ & 1000 & 15.3 & $ 5 $ & 3 & 0.1 \\   %BHNSq
			$ P_s $ & $ A_\varphi \propto \rho $ & 100 & 15.8 & $ 1.8 $ & 5 & 0.05\\   %BHNSs
			$ T_s $ & $ A_\theta \propto \rho $ & 1 & 16.5 & $ 4 $ & 4 & 4 \\ %BHNSts
                \hline
		\end{tabular}
		
		\caption{
			A summary of the models' parameters. The model names stand for hydrodynamic ($H$), poloidal ($P$), or toroidal ($T$) initial magnetic fields, with the subscripts indicate the strength of the field: zero ($0$), weak ($w$), canonical ($c$), or strong ($s$). $ A $ is the vector potential, $ \beta_p $ is the characteristic gas to magnetic pressure ratio, $ B $ is the characteristic initial magnetic field in the disk, $ t_f $ is the final time of the simulation with respect to the merger, $ M_{\rm ej} $ is the amount of unbound ejecta at the homologous phase, and $ t_b $ is the breakout time of the relativistic outflow from the disk winds.
    		}
    		\label{tab:models}
	\end{table}

In \textsc{h-amr}, we employ a relativistic gas equation of state, $p_g = (\gamma - 1) u_g$, where $p_g$ and $u_g$ are the gas pressure and internal energy densities, and $\gamma = 4/3$ is the adiabatic index. We emphasize that this equation of state and the lack of neutrino cooling may alter the outflows, particularly during the first few hundred ms, as we discuss later. The grid in spherical-polar coordinates is uniform in $\log r$, $\theta$ and $\varphi$, extending from $ r = r_g $ to $ r = 10^6\,r_g$, where $r_g=G\mbh/c^2$ is the BH gravitational radius. The base grid resolution is $N_r\times N_\theta\times N_\varphi = 384\times 96\times 96$ cells. Using static mesh refinement, we double the base resolution (quadruple in model $ T_s $) in all dimensions at $4 < r/r_g < 100 $. The higher resolution in the disk compared to the immediate vicinity of the BH allows our simulations to resolve the wavelength of the fastest-growing MRI mode \citep{Balbus1991}. We verify this by calculating the MRI quality factor $ Q_f $, which gives the number of cells per the MRI wavelength, and find that $ Q_f \gg 30 $ in all models, significantly higher than the $ Q_f \sim 10 $ required for resolving MRI \citep{Hawley2011}. We conduct a resolution convergence test in Appendix~\ref{sec:convergence}. We also employ three levels of AMR in the relativistic outflows, by requiring that both the jet and the cocoon are resolved at all radii by at least 96 cells each, based on their opening angle calculated using the magnetization criterion \citep[see details in][]{Gottlieb2022b}. Overall, at the highest refinement level, the effective resolution in the grid is $3072\times 768\times 768$ cells.
 
\section{Jet Launching}\label{sec:launching}

    \begin{figure}
    \centering
    	\includegraphics[scale=0.24]{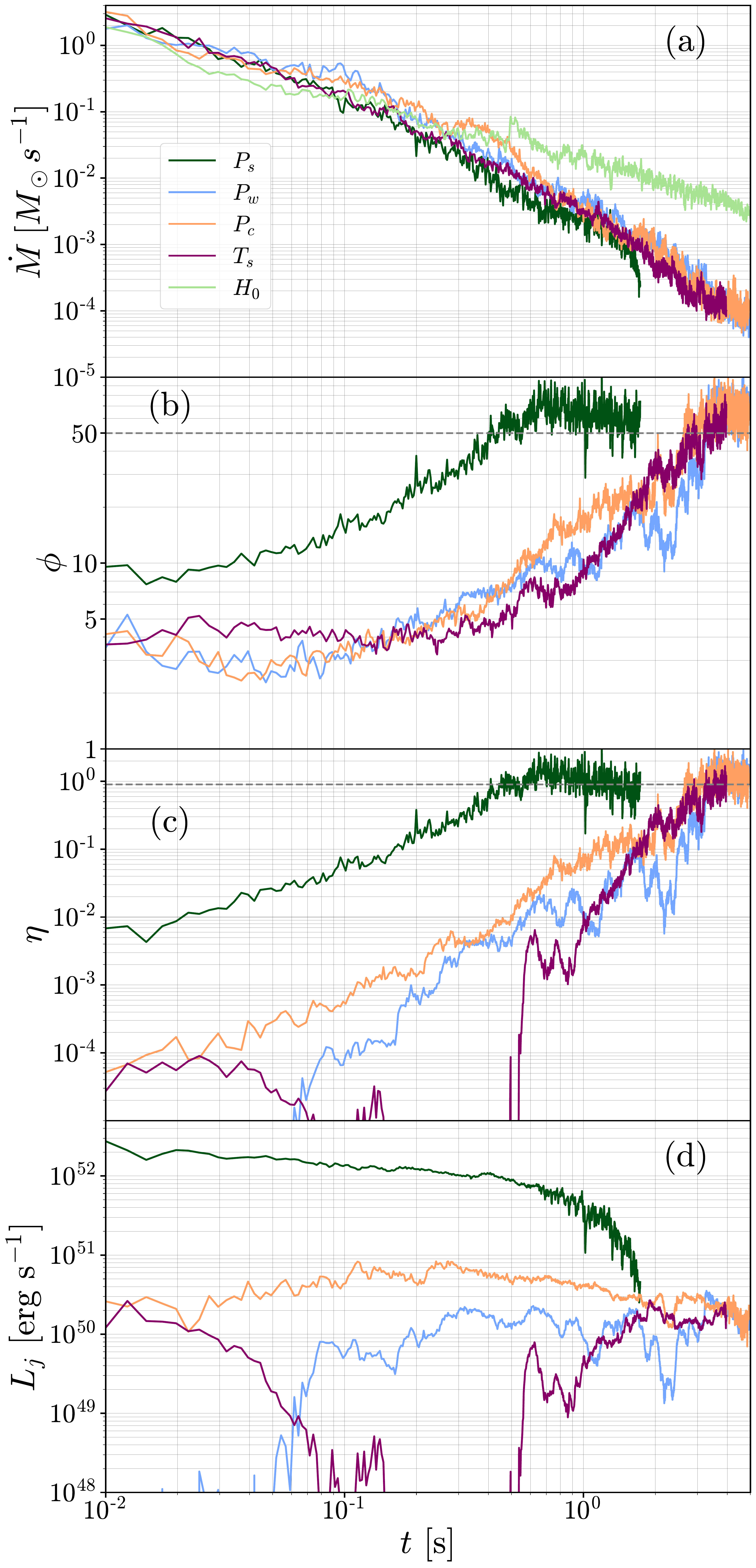}
     \caption{Time evolution at $ r = 5\,r_g $ for different models.
     {\bf Panel (a)}: Mass accretion rate in all models with a magnetized disk features $ \dot{M} \propto t^{-2} $, whereas in the hydrodynamic model $ H_0 $, $ \dot{M} \propto t^{-1} $.
     {\bf Panel (b)}: Dimensionless magnetic flux $ \phi = \Phi/\sqrt{\dot{M}cr_g^2} $, where $ \Phi $ is the magnetic flux. When the initial profile is a strong poloidal field, the flux reaches a MAD state within $ t \lesssim 1 $~s. When the initial magnetic field is weaker or purely toroidal, the magnetic flux accumulates slower on the BH horizon, and the MAD onset is delayed. Ultimately, all models turn MAD when $ \phi \approx 50 $ (dashed line).
     {\bf Panel (c)}: The jet launching efficiency increases gradually as more flux is threading the BH, until it reaches the maximum launching efficiency $ \eta \approx 0.9 $ in the MAD state (dashed line).
     {\bf Panel (d)}: The jet luminosity, $ L_j = \eta\dot{M}c^2 $, is roughly constant once the jet launching has been established by virtue of decreasing accretion rate and increasing efficiency. Once the system reaches the MAD state, the jet efficiency saturates, and the jet luminosity falls off proportional to the mass accretion rate.
    }
    \label{fig:launching}
    \end{figure}

Within a few milliseconds after the merger, an accretion disk with a nearly Keplerian rotation profile forms around the BH resulting from the merger. We investigate the impact of introducing different magnetic field configurations in the disk. In all models considered, the initial magnetic flux threading the BH at early times is insufficient to lead to a dynamically important magnetic flux and the magnetically arrested disk (MAD) state \citep{1974Ap&SS..28...45B,1976Ap&SS..42..401B,2003PASJ...55L..69N,2003ApJ...592.1042I,2008ApJ...677..317I,Tchekhovskoy2011}. During this early stage, accretion onto the BH is primarily driven by the MRI within the disk. As a result, the presence of magnetic fields enhances the accretion rate compared to unmagnetized disks, where only hydrodynamic instabilities are present. After a brief period of $ t \lesssim 10 $ ms, the availability of gas for accretion becomes more limited in magnetized disks. Consequently, the accretion rate declines rapidly, following a power-law of $ \dot{M} \propto t^{-2} $, as seen in Figure~\ref{fig:launching}(a), and consistent with simulations that included a neutrino leakage scheme \citep[e.g.,][]{Fernandez2015,Fernandez2017,Fernandez2019,Christie2019,Hayashi2022a}. Most of the accreted matter originates in the post-merger disk, which is $ \sim 0.8M_{\rm rem} $. The decline in the accretion rate for magnetized disks is faster than the case of a purely hydrodynamic disk, where $ \dot{M} \propto t^{-1} $, driven by shocks between spirals in the disk (see Appendix~\ref{sec:hydro_disk}).

Figures~\ref{fig:launching}(b,c) depict the progressive amplification of the dimensionless magnetic flux on the BH and of the jet launching efficiency owing to the reduction in the mass accretion rate. Once the dimensionless flux reaches saturation at $ \phi \approx 50 $, the disk transitions to a MAD state \citep{Tchekhovskoy2011,2015ASSL..414...45T}. Concurrently, the BH achieves its maximum jet launching efficiency of $ \eta \approx 0.9 $, consistent with the expected behavior for a BH spin of $ a = 0.86 $ \citep{Lowell2023}.

Fig.~\ref{fig:launching}(d) shows the jet power, defined as
\begin{equation}
    L_j = \int_{r_g}\sqrt{-g}(-T^r_t-\rho u^r)c^2\mathrm{d}\theta \mathrm{d}\varphi\,,
\end{equation}
and considering only fluid elements with $ \sigma > 1 $, where $ g $ is the metric determinant, $ T^r_t $ denotes the radial energy flux density expressed in terms of the stress-energy tensor $ T $, and $ u^\mu $ is the four-velocity such that $ \rho u^r $ represents the radial mass-energy flux density. The jet luminosity can also be expressed through the efficiency $\eta$ as $ L_j = \eta\dot{M}c^2 $. As the accretion rate decreases and the jet efficiency increases, the jet power remains relatively constant: this is primarily due to the approximate constancy of the large-scale vertical BH magnetic flux, which controls the jet power. The first model to reach a MAD state is $ P_s $, where the initial magnetic field corresponds to the strongest poloidal field, and the disk becomes MAD less than a second after the merger. Eventually, the dimensionless magnetic flux, $\phi$, and jet efficiency, $ \eta $, reach and remain saturated at their asymptotic MAD values. As a result, the jet power begins to decay, following a power-law of $ L_j \propto \dot{M} \propto t^{-2} $. In our simulations with magnetic fields, all disks eventually reach a MAD state within several seconds, marking the end of the jet \citep[see][who explored a similar idea in the context of long GRBs]{Tchekhovskoy2015}. Disks with weaker initial magnetic fields take longer to reach the MAD state compared to those with stronger initial fields \citep[see also][]{Christie2019,Fernandez2019}. Of particular interest is the simulation with initial toroidal field $ T_s $, which requires efficient dynamo process to form a global poloidal field and generate relativistic outflows. The ongoing dynamo process also results in higher variability in the jet launching efficiency compared to the initially poloidal field configurations. Consequently, although its initial $ \beta_p $ is the highest, the jet forms last compared to the initial poloidal configurations, and has to punch through more massive winds.

Interestingly, putting the above another way, we find that either the jet is too luminous (in model $ P_s $), or its launching process is too long (in all other models) compared to typical sGRBs. In order for jets to be consistent with the observed jet power, they cannot be too luminous, i.e., they have to reach maximum efficiency only after the accretion rate drops substantially from $ \dot{M} \sim \msun\,{\rm s}^{-1} $. However, this necessarily requires the jet launching to be longer than a typical sGRB duration of $\lesssim 1\,{\rm s}$. We stress that the exclusion of the alpha recombination effect in our simulations is unlikely to impact this conflict. The reason is that these factors only start to influence the mass accretion rate after the neutrino luminosity decreases at $ t \sim 0.5\,{\rm s}$, shifting the accretion rate from $ \dot{M} \sim t^{-2} $ to $ \dot{M} \sim t^{-3} $ \citep{Haddadi2023}. As a result, their impact on the jet luminosity is negligible within the typical duration of sGRBs, $ t \lesssim 1\,{\rm s}$.

A similar challenge of excessive jet power was recently found for long GRBs in collapsars and can be resolved by the requirement that the BH is slowly spinning, such that the jet power is reduced to reduce the tension with that of long GRBs \citep{Gottlieb2023a}. However, in BH--NS mergers, low spin BHs cannot resolve the problem. The reason is that low post-merger BH spins require low angular momentum in the merger, such that the NS is not disrupted to form a post-merger disk, and hence there are no jets. However, a low pre-merger BH spin (which would increase after the merger) or a higher mass ratio would reduce the disk mass to $ \sim 10^{-3}\,\msun $ and subsequently the accretion rate. If the post-merger evolution scales with the disk mass, model $ P_s $ would yield a jet with a typical sGRB power and duration. Similar but weaker effects can also be caused by a tilt angle of the binary orbit. Nonetheless, higher BH spins, such as the one in this study, may provide a natural explanation for the origin of the recent detection of sGRB 211211A \citep[e.g.,][]{Rastinejad2022,Troja2022,Yang2022}. This event featured bright kilonova emission, as expected from the large debris of high BH spins, and a $ \sim 10 $~s burst. We leave a full investigation of the GRB characteristics as a function of the magnetic field configuration to follow-up work.

Our results exhibit similarities to those presented in \citet{Hayashi2022a,Hayashi2023}, where the magnetic field evolved self-consistently from the disrupted NS. This implies that the observed behavior is not a consequence of our choice of the post-merger magnetic field. Once the jet forms within $ \sim 1 $~s, it maintains a relatively constant jet luminosity, indicating an increasing efficiency. \citet{Hayashi2023} reported that the disk does not reach a MAD state, and their figure 17 demonstrates that the dimensionless magnetic flux has not yet reached saturation at a value of $ \phi \approx 50 $. A longer integration can lead to the attainment of a MAD state, at which point a decline in jet luminosity will follow, as observed in our study. Initial toroidal fields necessitate stronger magnetization to generate sufficiently robust poloidal flux through the dynamo process early on \citep{Christie2019}. This may also elucidate why \citet{Ruiz2018,Most2021,Gottlieb2022d} did not find evidence of relativistic jets when the initial magnetic flux has a toroidal configuration with $ \beta_p \gtrsim 10 $, as the time over which a poloidal field is generated will be longer than the simulation time for such relatively high $ \beta_p $ values.
 
\section{Sub and Mildly Relativistic Outflows}\label{sec:ejecta}

\begin{figure}
    \centering
    	\includegraphics[width=3.3in]{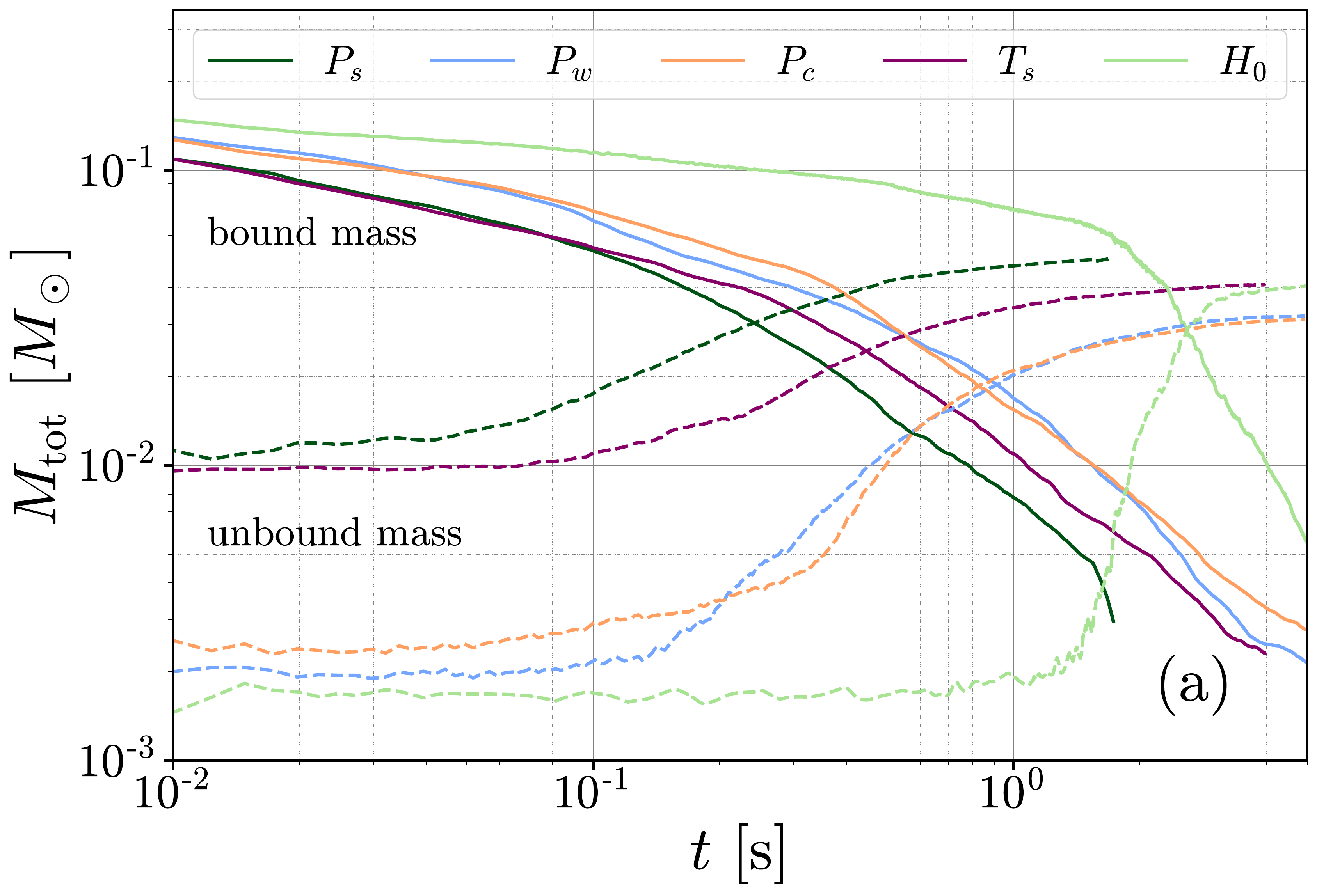}\\
    	\includegraphics[width=3.3in]{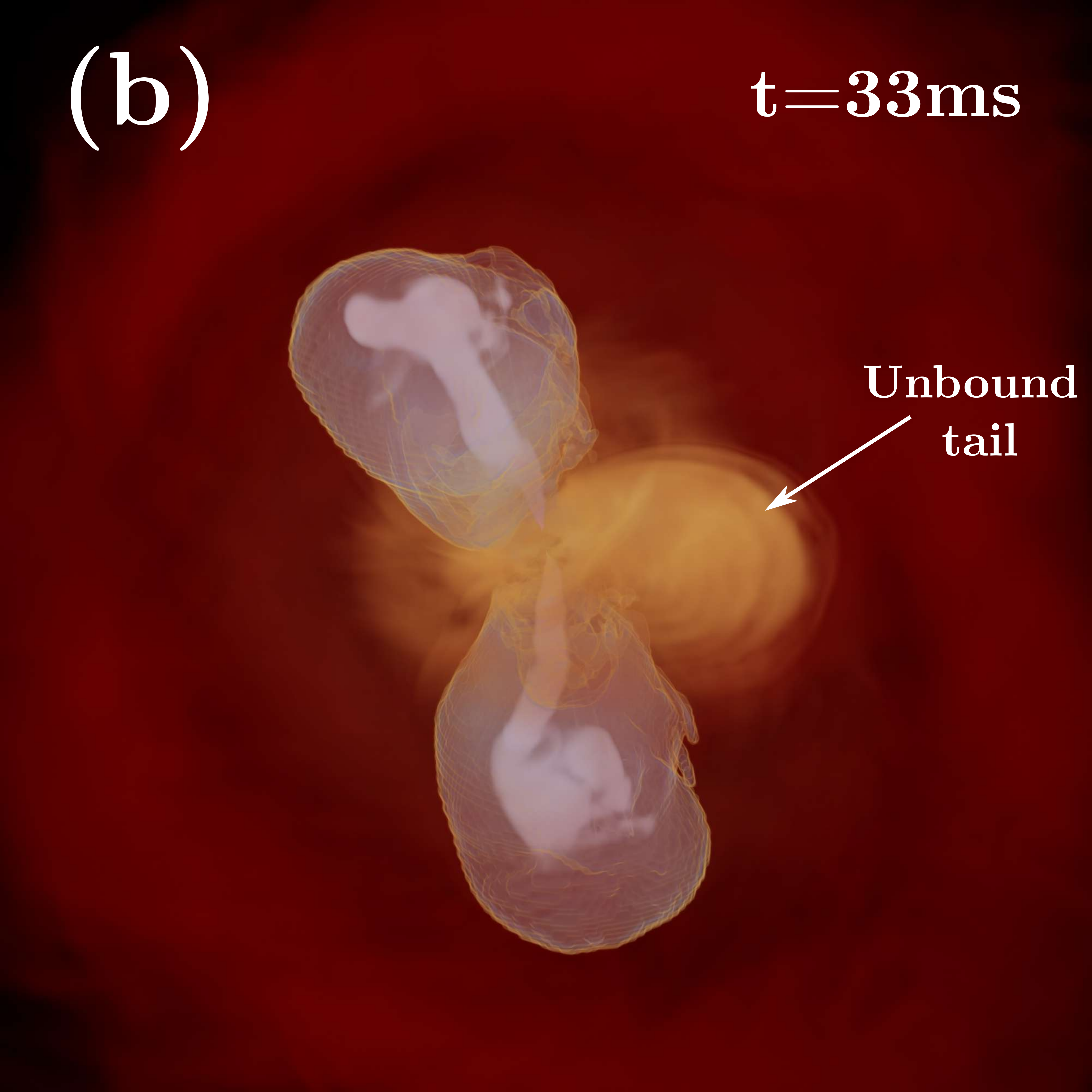}
     \caption{
     {\bf Panel (a)}: Time evolution of the bound (solid lines) and unbound (dashed lines) ejecta in the different models. When the disk is magnetized, MRI-driven winds and the extended relativistic outflow structure unbind a fraction of the merger debris. Stronger fields lead to earlier and stronger outflows.
     {\bf Panel (b)}: 3D rendering of the system at 33 ms post-merger in simulation $ P_c $, with colors reflecting the mass density for non-relativistic gas and proper velocity for relativistic gas (spin axis direction is up and the jet--cocoon outflow extends to $ r \approx 2000 $ km). The jets (white) are launched from the BH into the expanding disk winds (red), generating a cocoon (light-blue). The bound tidal tail becomes part of the accretion disk whereas the unbound tail, shown as the extended yellow component on the right, expands along the equatorial plane.
     }
    	\label{fig:bound}
    \end{figure}

Figure~\ref{fig:bound}(a) illustrates the quantities of bound (solid lines) and unbound (dashed lines) mass, as determined by the Bernoulli parameter criterion $ -(h+\sigma)u_t>1 $, where $ h = 1+4p_g/ \rho c^2 $ is the specific enthalpy, and $u_t$ is the covariant time-component of the four-velocity vector. In comparison to our hydrodynamic model $H_0$, all of our models incorporating magnetic fields exhibit stronger outflows that emerge early on due to the combined effects of the MRI in the disk and strong relativistic outflows, with stronger magnetic fields resulting in earlier and more substantial ejections of mass. However, we note that the difference in the final amount of unbound mass is comparable to the difference at the onset of the GRMHD simulation between strongly and weakly magnetized disks \citep[see also][]{Hayashi2022a}. This implies that introducing strong initial magnetic fields induces a firm relaxation and unbinds the highly magnetized fluid promptly. Consequently, without further investigation, we cannot definitively confirm whether the variations in the unbound mass are physical. In unmagnetized disks, the low accretion rate sustains a heavier disk, which may explain the ejection of more massive outflows compared to weakly magnetized disks, as shown by the asymptotic behavior of the unbound ejecta in Fig.~\ref{fig:bound}(a). Similar to accretion, the origin of the outflows in model $ H_0 $ is the shocks generated by the interaction between spiral density waves in the disk (see Appendix~\ref{sec:hydro_disk}).

    \begin{figure*}
    \centering
    	\includegraphics[width=1.75in]{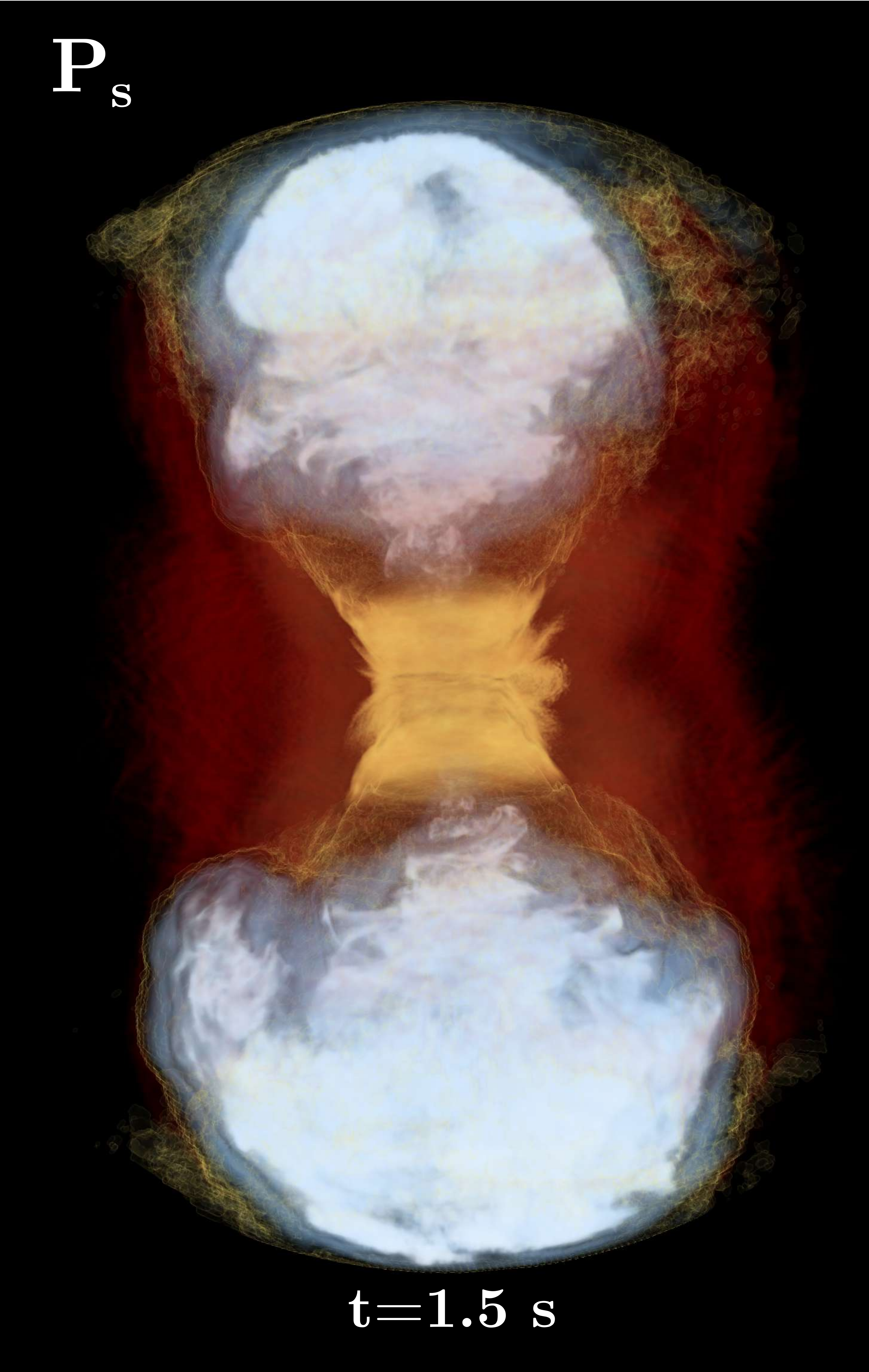}
    	\includegraphics[width=1.75in]{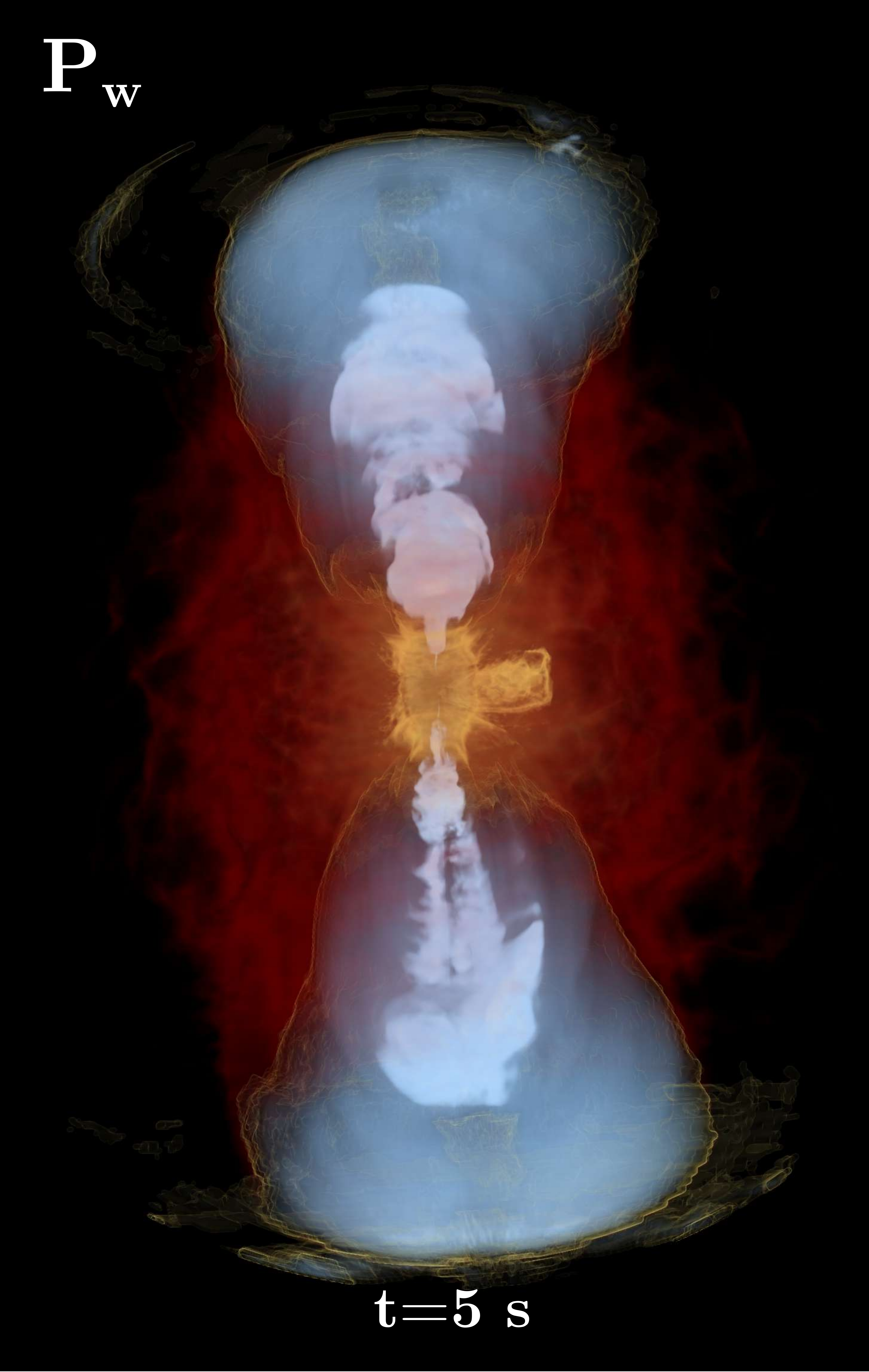}
    	\includegraphics[width=1.75in]{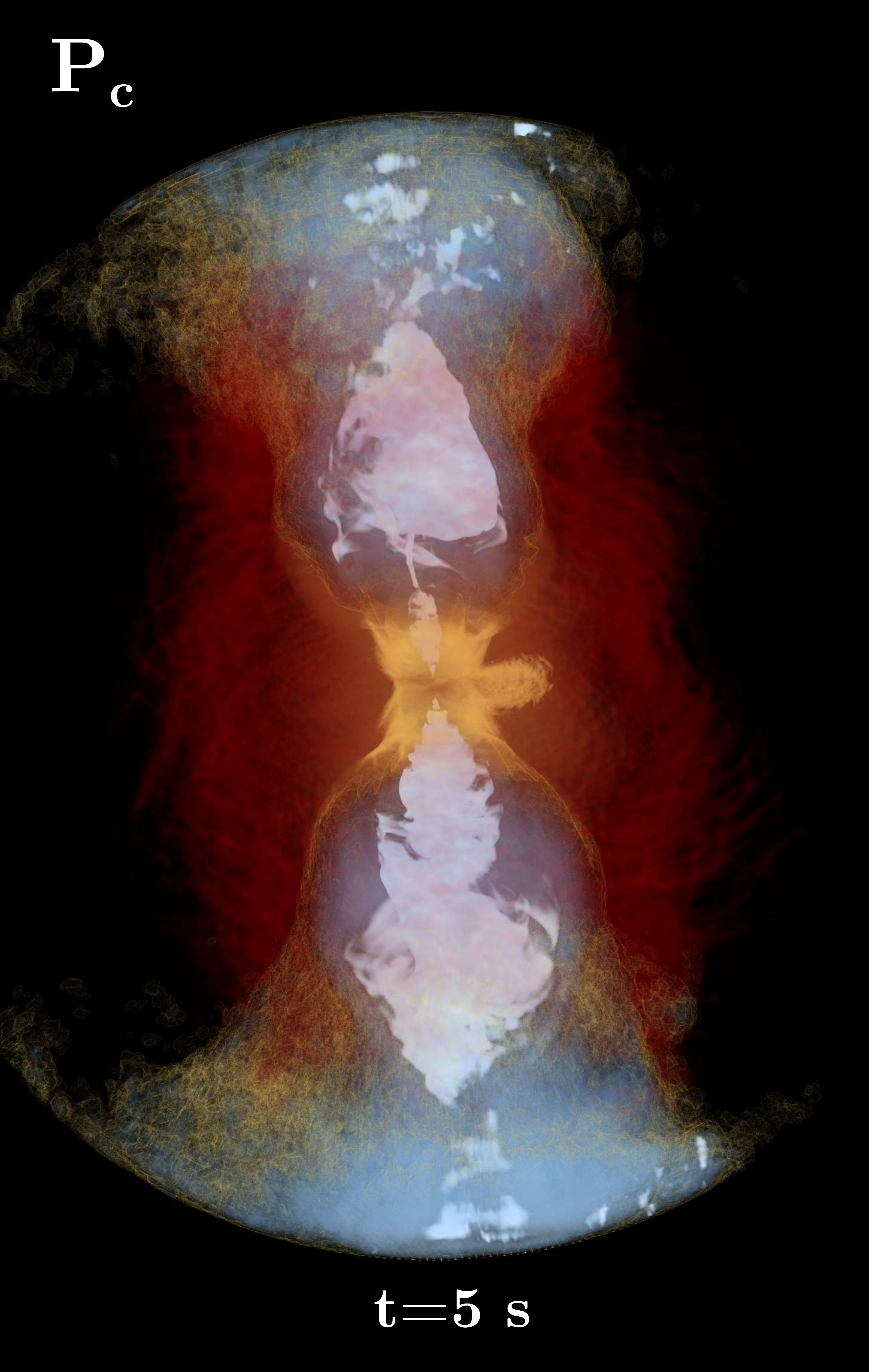}
    	\includegraphics[width=1.75in]{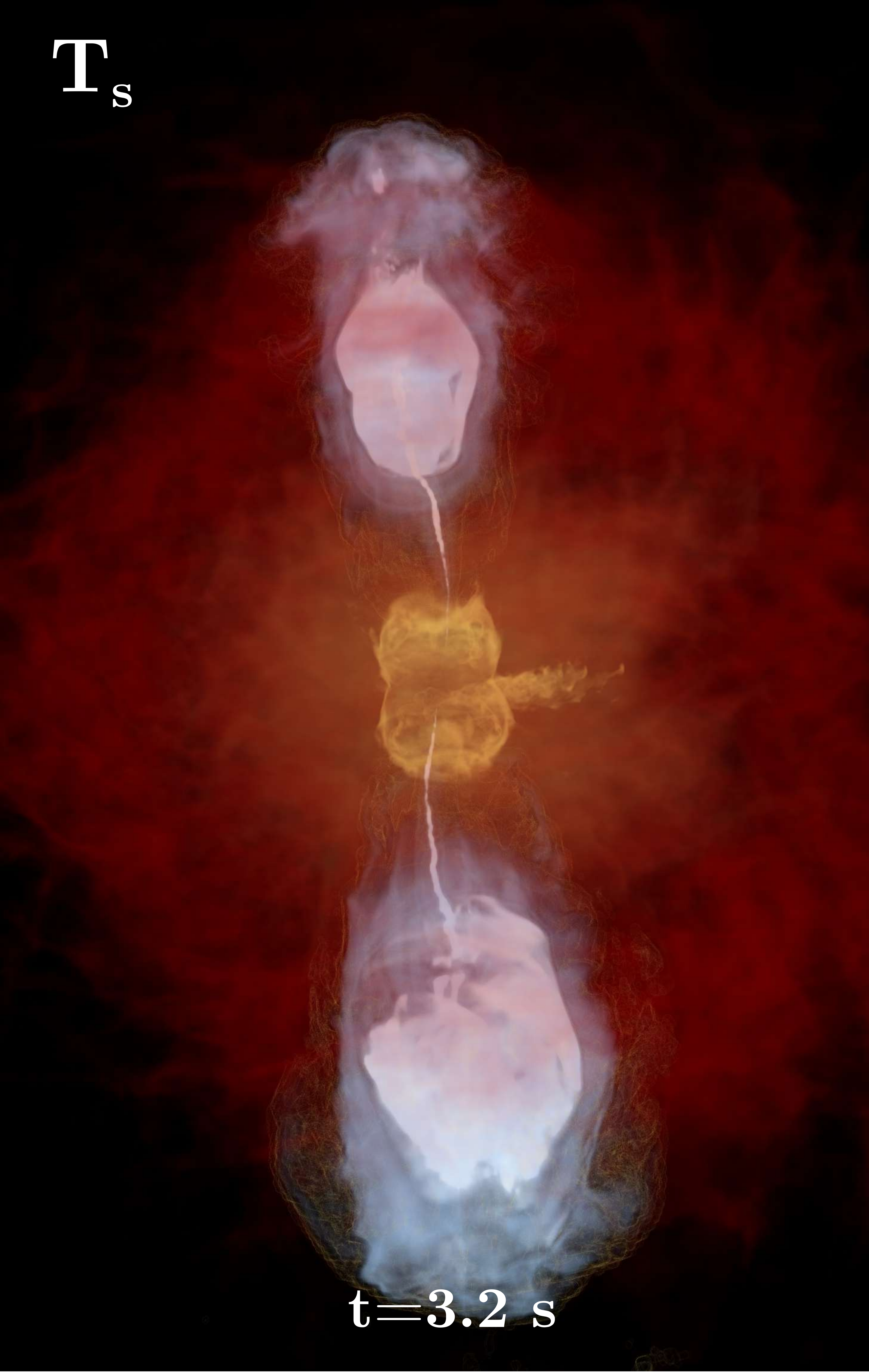}
     \caption{Volume rendering of the final snapshot of each simulation (see Table~\ref{tab:models}). The outflows in the poloidal field configurations reach $ r \approx ct $ whereas in model $ T_s $ the outflow front is at $ r \approx 6.4\times 10^{10} $ cm. The BH spin axis direction is up, and different colors show different quantities. The asymptotic proper velocity is depicted in blue, revealing the presence of a sub- and mildly relativistic cocoon (dark) surrounding relativistic jets (light). Red and yellow components represent the mass density, showing the merger ejecta and the accretion disk components, respectively. See full movies of models $ T_s $ and $ P_c $ at \url{http://www.oregottlieb.com/bhns.html}.
     }
     \label{fig:maps}
    \end{figure*}

The ejecta mass in all models reaches its asymptotic value by the end of the simulations, $ \sim 3--5\times 10^{-2}\,\msun $, as indicated by the asymptotic behavior of the unbound mass. Of particular interest is the fate of the tidal tail. Figure~\ref{fig:bound}(b) shows a 3D rendering at small distance from the BH in model $ P_c $, demonstrating that the tidal tail is composed of bound and unbound (extended yellow component on the right) components. All models feature a similar behavior of the tidal tail. The innermost part of the bound component merges with the disk at early times, slightly increasing the disk mass. The unbound component ultimately influences the kilonova composition \citep[see e.g.,][]{Fernandez2017}. We find no evidence for late fallback accretion, as all models exhibit a smooth power-law decline in the mass accretion rate (Fig.~\ref{fig:launching}(a)). However, in the absence of neutrino cooling, heavy nuclei, and radioactive heating \citep[e.g.,][]{Desai2019,Haddadi2023}, we cannot preclude a definitive conclusion regarding late-time fallback of the tail. In a recent study, \citet{Metzger2021} found that marginally bound tail progressively falls back into the disk, but it quickly becomes unbound due to its inability to efficiently cool down. If indeed the bound tail fails to reach the BH, hindering the increase in mass accretion rate, the activation of the sGRB-EE by the tidal tail \citep{Metzger2010} is improbable.

Figure~\ref{fig:bound}(b) also portrays the preeminent initiation of disk winds compared to the emergence of the relativistic jets. This indicates that the trajectory of the jet intersects with the disk winds, resulting in the formation of a layer of shocked cocoon. Consequently, the presence of dynamical ejecta is not a prerequisite for cocoon formation, suggesting that cocoons accompany all jets \citep[see also][]{Gottlieb2022d}. The emission from the cocoon spreads across wide viewing angles and could have a noteworthy impact on BH--NS mergers, similar to its influence in the BNS merger GW170817. In a companion paper \citet{Gottlieb2023b}, we perform a detailed calculation of the early near-UV/optical emission. We find that the cocoon generates a bright signal, exhibiting an absolute magnitude of $M_{\rm AB} \approx -15 $ for a few hours after the merger. In future work, we will investigate the effect of neutrino cooling on the timing and amount of disk wind mass ejecta.

\section{Relativistic outflows}\label{sec:outflow}

    \begin{figure}
    \centering
    	\includegraphics[scale=0.24]{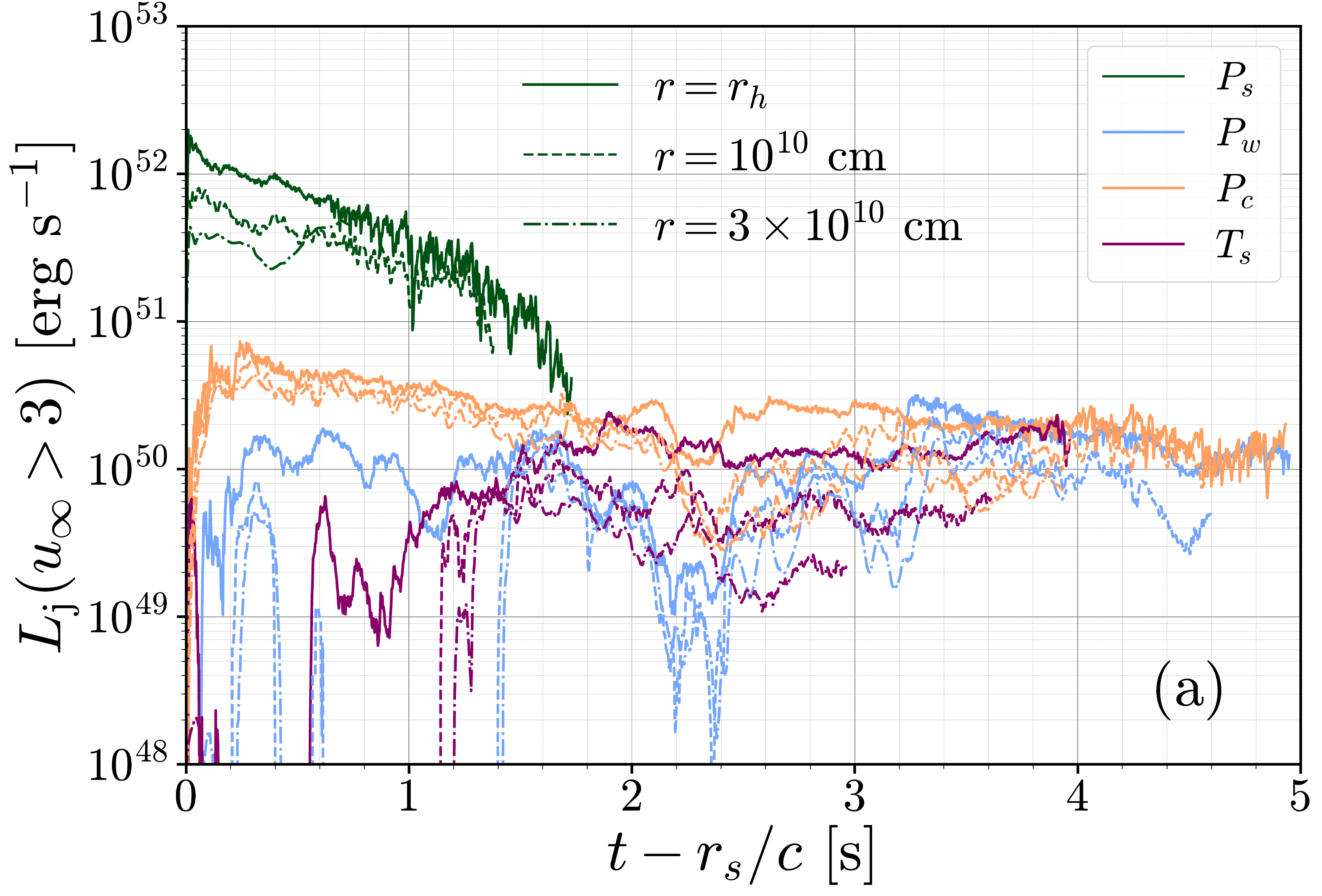}
    	\includegraphics[scale=0.235]{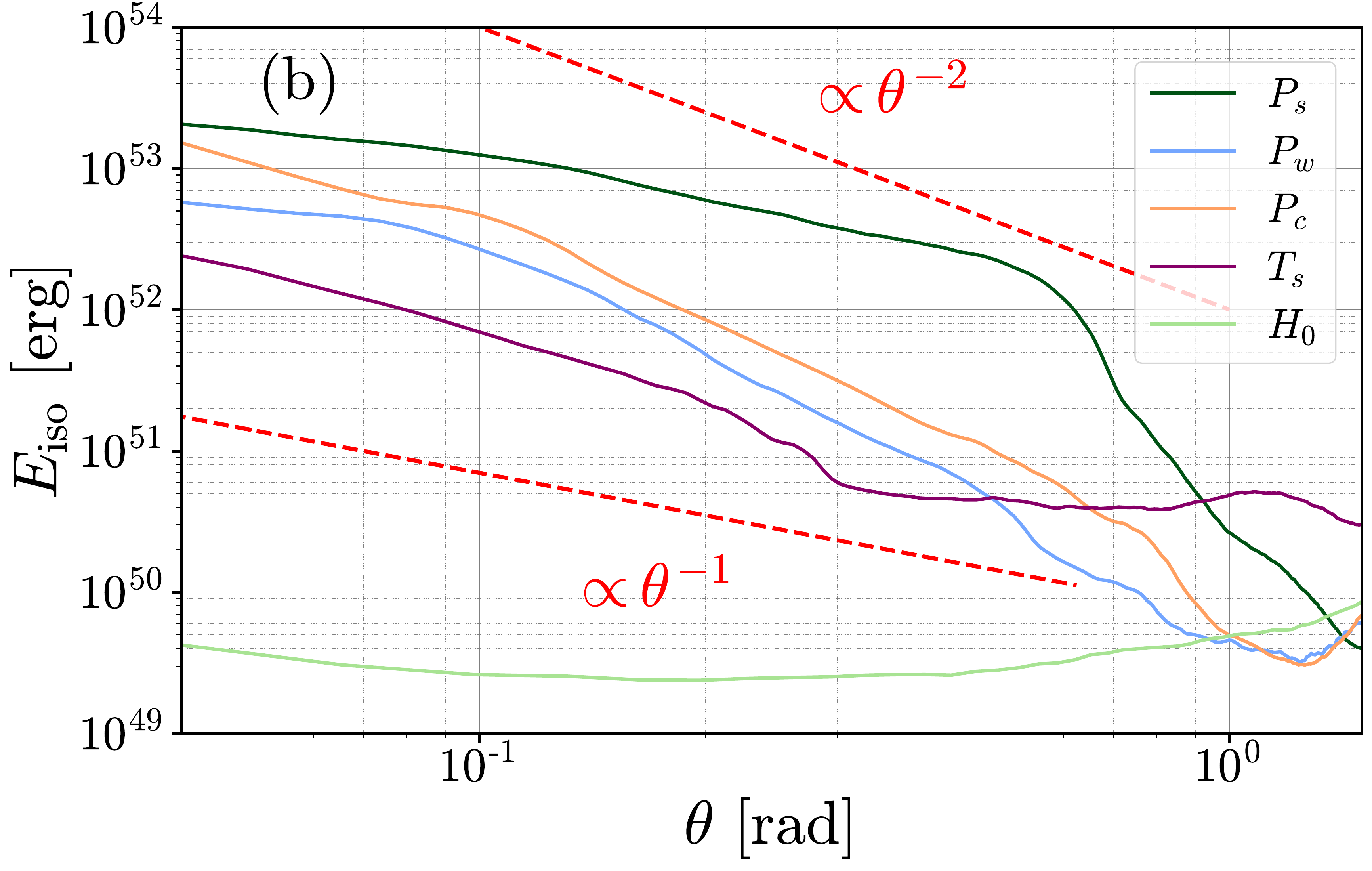}
            \includegraphics[scale=0.235]{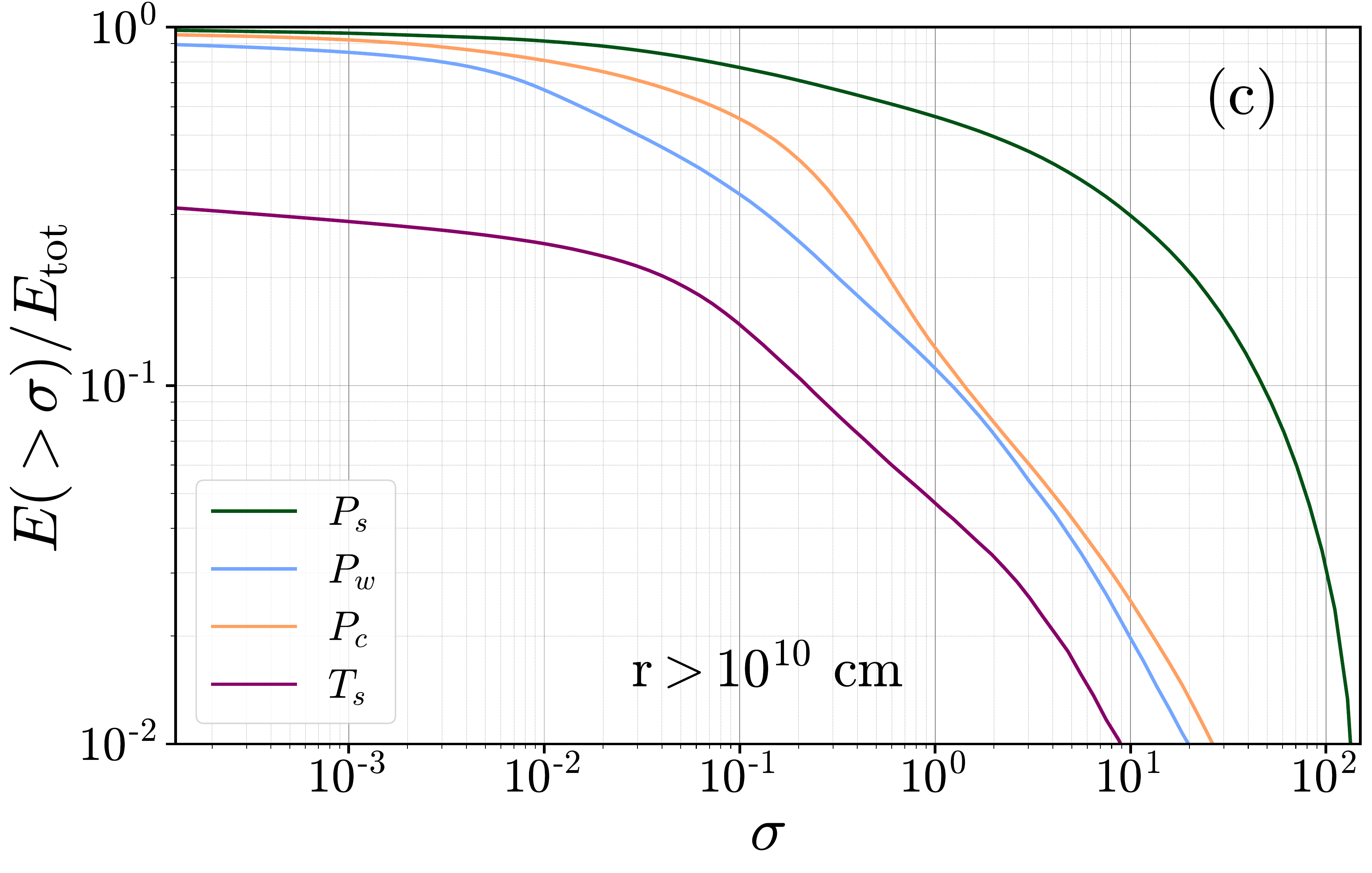}
     \caption{
     {\bf Panel (a)}: Time evolution of the energy contained in the material with asymptotic proper velocity $u_\infty > 3$ integrated on the horizon, $ r_s = r_{\rm H} $ (solid lines), at $ r_s = 10^{10}\,{\rm cm} $ (dashed lines), and $r_s = 3\times10^{10}\,{\rm cm} $ (dash-dotted lines). The time axis is shifted by the time it takes ultra-relativistic fluid elements to reach $ r_s $, i.e. $ t-r_s/c $. Except for the initial launching, most jets retain their power profile throughout their propagation. In model $ T_s $ the jets are subject to strong mixing and thus their power drops with radius.
     {\bf Panel (b)}: Isotropic equivalent energy of the homologously expanding gas, at $ r \gtrsim 10^{10}\,{\rm cm} $ for magnetized outflows and $ r > 10^9\,{\rm cm} $ for the pure hydrodynamic model $ H_0 $. In the absence of relativistic outflows in model $ H_0 $, the post-merger evolution results in quasi-isotropic winds. The initial toroidal field takes longer to launch a steady jet, which results in a wider outflow structure. The jet in model $ P_s $ is too powerful to be collimated by the winds.
     {\bf Panel (c)}: Cumulative energy in plasma at $ r > 10^{10}\,{\rm cm} $ with magnetization larger than $ \sigma $, normalized by the total energy. All jets contain at least a few percent of their total energy at high magnetization, $ \sigma \gg 1 $, suggesting that magnetic processes may play a key role in the sGRB prompt emission.
     }
    	\label{fig:jet_structure}
    \end{figure}
    
Figure~\ref{fig:maps} displays the final snapshot of the simulations, capturing a 3D rendering of the four models characterized by magnetic fields that generate relativistic outflows. All models exhibit disk winds (red-yellow), which are shocked by the relativistic jets (light blue) to form a hot cocoon (dark blue). We find that all jets wobble, similar to collapsar jets \citep{Gottlieb2022b}, but to a lesser extent with a tilt angle of $ \theta_t \sim 5^\circ\mbox{--}10^\circ $. The jet opening angle increases over time due to weaker collimation by the disk winds \citep[see also][]{Hayashi2022a,Hayashi2023}. In all simulations, the jet head is relativistic, so that its evolution follows the solution of jet propagation in the static media regime \citep{Gottlieb2022f}.

Figure~\ref{fig:jet_structure}(a) shows the evolution of the jet power as a function of the distance from the BH. The solid lines represent the jet power on the horizon, $ r_s = r_{\rm H} = r_g(1+\sqrt{1-a^2}) $, where $ a $ is the BH spin, which is identical to Fig.~\ref{fig:launching}(d). The dashed (dash-dotted) lines depict the jet power at $ r_s = 10^{10}\,{\rm cm} $ ($ r_s = 3\times 10^{10}\,{\rm cm} $). To facilitate comparison, the $ x $-axis represents the retarded time $ t - r_s/c $, ensuring that ultra-relativistic elements appear at the same time. Initially, the jet undergoes stronger interactions with the disk winds, leading to a reduction in jet power with increasing $ r $. This effect is particularly pronounced in model $ T_s $, where the early-time jets are weak due to the absence of a global poloidal field. For initial poloidal configuration, it shows that as time progresses, the interaction between the jet and the winds weakens, allowing the jets to retain their power as they propagate. This result suggests that for such jets, the power on the horizon may provide a good indication of the time evolution of the jet power close to the emission zone. However, it is still necessary to consider the emission mechanism and radiative efficiency to draw conclusions about the GRB emission.

For the toroidal initial magnetic field configuration, model $ T_s $, there is an initial jet launching owing to random poloidal loops that are generated by the dynamo process. Magnetic reconnection leads to a quiet episode before the launch of a steady jet when the dynamo process generates a sufficiently strong global poloidal field, as seen in Fig.~\ref{fig:launching}(d). This picture is similar to the BH model suggested by \citet{Kisaka2015}, however here the initial jet is too weak to survive the interaction with the disk winds. By the time the steady jet is launched, it needs to interact with more massive ejecta for a longer duration, leading to a loss of jet power and greater deposition of energy into the cocoon. In fact, when the jet is launched, most of the ejecta is already positioned along its trajectory, resulting in the jet encountering an isotropic equivalent mass of $ \sim 5\times 10^{-2}\,\msun $. The jet ultimately escapes from the ejecta after several seconds, consistent with the theoretical predictions for typical sGRB power in such ejecta \citep{Gottlieb2021a,Gottlieb2022f}. If typical sGRB jets break out within $ \sim 1 $ second \citep{Moharana2017}, this may necessitate some poloidal component to already be present at the time of the merger in order to produce typical sGRBs. On the other hand, such jets can explain long-duration sGRBs such as GRB 211211A. Furthermore, if a long delay between the GW signal and the GRB prompt emission is observed, this may imply that the post-merger magnetic field is predominantly toroidal.

Figure~\ref{fig:jet_structure}(b) depicts the angular distributions of the isotropic equivalent energy in different models, defined as
\begin{equation}
    E_{\rm iso}(\theta) = 2\frac{\mathrm{d}}{\mathrm{d}{\rm cos}\theta}\int_0^\infty\int_0^\theta\int_0^{2\varphi}\sqrt{-g}(-T^t_t - \rho u^t)c^2r^2{\rm sin}\theta\mathrm{d}r\mathrm{d}\theta'\mathrm{d}\varphi.
\end{equation}
In the absence of (or for negligible) magnetic fields in the disk (model $ H_0 $), the outflow exhibits a quasi-isotropic distribution with a slight excess of energy along the equatorial plane due to disk winds. The introduction of large-scale magnetic fields triggers the formation of relativistic jets, causing a redistribution of the gas. In our canonical model $ P_c $ and a similar field configuration in model $ P_w $, the energy decreases away from the polar axis, following a power-law $ E_{\rm iso} \sim \theta^{-2} $. This structure features more energy in the cocoon compared to the one emerging when an idealized torus is assumed \citep[model $ V $ in][]{Gottlieb2022d}, or purely hydrodynamic and weakly magnetized jets \citep{Gottlieb2020b,Gottlieb2021a}. We find modifications in this structure when strong initial fields are present. In model $ P_s $, the jet power significantly exceeds the rest mass energy of the local ejecta, resulting in a weakly collimated jet, as can also be seen in Fig.~\ref{fig:maps}. Consequently, the energy distribution follows $ E_{\rm iso} \sim \theta^{-1} $ up to the characteristic angle of the relativistic outflows at $ \theta \approx 0.5\,{\rm rad} $. In model $ T_s $, the energy deposition of the jet into the cocoon results in a wider distribution of energy at larger angles at the expense of energy along the pole.

Figure~\ref{fig:jet_structure}(c) presents the cumulative magnetic energy fraction out of the total energy. We show here for the first time that sGRB jets maintain a significant reservoir of magnetic energy far from the launching point, with $ E_{\rm iso} \sim 10^{50} \mbox{--} 10^{52}\,{\rm erg} $\footnote{Unfortunately, it is challenging to verify that physical processes can fully account for the magnetic dissipation into heat, such that there is no further contribution by numerical dissipation. Nevertheless, collapsar simulations have shown that jets are subject to stronger dissipation owing to denser environment, compared to our findings \citep{Gottlieb2022b}, indicating that the dissipation is at least partly physical. Overall, we consider the dissipation seen in our simulations to be an upper limit (lower limit on the resultant magnetization).
}. Such substantial magnetic energy implies that magnetic processes, such as magnetic reconnection and/or synchrotron emission, play a crucial role in the emission of sGRBs. This finding is novel, as previous first-principles simulations of jets in binary mergers have not considered the launch of highly magnetized jets with $ \sigma_0 > 100 $. While \citet{Gottlieb2022b} investigated such highly magnetized jets in the context of collapsars, the dense stellar envelope in those scenarios led to significantly stronger interactions between the jet and the medium, resulting in a low asymptotic magnetization of $ \sigma \sim 0.1 $.

\section{Conclusions}\label{sec:conclusions}

We performed the first BH--NS simulations that track the merger evolution from the pre-merger phase to the homologous expansion of the outflows at distances $ r > 10^{11}\,{\rm cm} $. To achieve this, we employed a technique where we remapped numerical relativity simulations to 3D GRMHD simulations at 8 ms after the merger, as the metric does not undergo significant changes beyond that point. The properties of the binary system, including a mass ratio of $ q = 2 $, a high post-merger BH spin of $ a = 0.86 $, and an aligned spin-orbit configuration, favor the presence of a large mass reservoir outside the ISCO, facilitating efficient jet launching. We investigated the formation of outflows for various configurations of magnetic fields in the post-merger accretion disk.

We found that in all configurations, the large-scale vertical magnetic flux accumulated on the BH does not become dynamically important right away and, hence, the accretion disk does not immediately enter a MAD state after the merger. Instead, the magnetic flux quickly accumulates and remains approximately constant on the BH thereafter,  $\Phi\approx \text{constant}$. As $\dot M$ decreases, the dimensionless magnetic flux $\phi\propto \Phi/\dot{M}^{1/2}$ gradually grows on the BH, leading to an increase in jet efficiency, $\eta = L_j/\dot M c^2 \propto \phi^2$. Simultaneously, the constancy of $\Phi$ results in a constant jet power, $L_j \propto \Phi^2 \approx \text{constant}$. Stronger initial fields lead to more luminous jets and reach MAD onset faster. Eventually, all disks reach the MAD state within several seconds, at which point the magnetic flux becomes too strong to all stay on the BH, and both $\Phi \propto \dot M^{1/2}$ and the jet power $L_j \propto \Phi^2 \propto \dot M$ begin to drop, such that the dimensionless magnetic flux and jet efficiency remain at the MAD level, $\Phi/\dot M^{1/2} \propto \phi \approx 50$ and $\eta = L_j/\dot M c^2 \approx 1$, respectively. Thus, the onset of the MAD state marks the end of a sGRB jet. This picture can be tested observationally as it has a clear observational signature at the end of the sGRB signal, $ L_j \sim \dot{M} \sim t^{-2} $.

While it has been speculated that BH--NS mergers produce less ejecta outside of the equatorial plane compared to BNS mergers, our simulations demonstrate that a significant amount, $ \sim 20\%-30\% $ of the total baryonic mass outside of the ISCO at the time of the merger, $ \sim 3--5 \times 10^{-2}\,\msun $, is ejected after the merger due to heating from disk turbulence. Stronger magnetic fields have the capability to unbind a larger amount of merger debris through stronger MRI heating, resulting in a greater reservoir of ejecta. It should be noted that this result is specific for our configuration, which produces a massive torus similar to that in BNS mergers. The disk mass is anticipated to be substantially smaller for lower BH spins or higher binary mass ratios, and can also be reduced by neutrino cooling. Since the cocoon is generated during the jet-wind interaction, a less massive disk would lead to a weaker wind and lighter cocoon. However, the angular structure of the cocoon seems to be independent of the cocoon energy \citep[see for comparison][]{Gottlieb2020b,Gottlieb2021a,Gottlieb2022d}. In the future, we will investigate how neutrino cooling affects the mass of the disk wind ejecta.

While revealing a notable interaction between the jet and polar disk winds, the simulations show that the jet elements manage to preserve the majority of their energy while propagating away from the BH. The stability of the jet is further demonstrated by the presence of a large reservoir of magnetic energy far from the BH, $ r > 10^{10}\,{\rm cm}$. Using an initially ultra-high magnetization of $ \sigma_0 = 150 $, we showed for the first time that at least a few percent of the total jet energy exists at $ \sigma \gg 1 $, highlighting the significance of magnetic processes in the sGRB prompt emission mechanism. Future work will investigate the implications of this result for the emission mechanisms in greater detail. Despite the jet successfully retaining its energy, the interaction of the jet with the disk winds inevitably inflates a hot cocoon, implying that cocoon formation is not conditional to the presence of dynamical ejecta. The jet--cocoon structure features an angular profile of the outflow isotropic equivalent energy that is consistent with a power-law distribution. The cocoon may have a crucial role in the detection of electromagnetic counterparts in BH--NS mergers, similar to its significance in BNS mergers. In a companion paper \citet{Gottlieb2023b}, we present evidence that the cocoon generates a remarkably bright near-UV/optical signal a few hours after the merger.

In cases where the magnetic field configuration in the post-merger disk is purely toroidal, the dynamo process generates a poloidal field and operates for $ \sim 1 $~s before launching a steady jet, depending on the field strength. During this initial period, only weak jets emerge from the BH, encountering increasing amounts of polar ejecta generated by disk winds. By the time the steady jet is launched, it becomes heavily contaminated by baryonic matter from the polar ejecta. As a result, the jet loses a significant portion of energy, which is redistributed at wider angles. Eventually, after a few seconds, the jet breaks free from the surrounding ejecta. Consequently, two intriguing possibilities arise: (i) Observation of a long delay between the GW signal and the GRB may indicate that the post-merger disk magnetic field is primarily toroidal, and (ii) A toroidal field configuration may provide some clues regarding the origin of EE: the MAD onset time is inversely proportional to the magnetic field strength. Thus, lowering the magnetic field by an order-of-magnitude compared to the toroidal model $ T_s $ may give rise to an early jet launch followed by a long ($ \sim 50 $ s) jet launching duration, which might be linked to the sGRB-EE mechanism.

Interestingly, we found that all jets display either excessive luminosity or too long a duration in comparison to typical sGRBs. The underlying reason can be explained as follows: in order for the jets to achieve typical sGRB luminosity, the maximum efficiency needs to be obtained after a significant decrease in the accretion rate. However, achieving this entails a longer duration for the jet launching process than what is typically observed in sGRBs. This behavior is not seen in simulations where the initial conditions are an analytic torus, presumably due to the exclusion of the violent merger, motivating the need for self-consistent simulations from the pre-merger phase. While such long sGRBs may explain the origin of the kilonova-associated long-duration sGRB GRB 211211A, they also introduce a fundamental challenge to our understanding of typical jet formation in all types of binary mergers. This could be alleviated by the requirement that the pre-merger BHs possess moderate spins of $ a \gtrsim 0.2 $ as suggested by LVK, or a higher binary mass ratio, as suggested by population synthesis models \citep{Belczynski2008}. In both cases, the merger debris might be sufficient to form a lighter disk of $ \sim 10^{-3}\,\msun $, where the accretion rate, and thus the jet power, would be substantially lower than the configuration studied here and may better agree with the observed sGRB luminosity and duration.

In a follow-up study, we will tackle the intriguing questions mentioned above about the relationship between the merger types and different characteristics of sGRBs. Our future research will investigate relativistic outflows, going beyond the specific BH--NS merger setup examined in this study. Specifically, we aim to examine the emergence of outflows and their prompt emission and EE in various BNS and BH--NS merger configurations, such as different mass ratios and BH spins. By conducting these models, we can acquire a comprehensive understanding of the possibilities for sGRBs in diverse configurations of binary mergers.

\begin{acknowledgements}

OG is supported by a CIERA Postdoctoral Fellowship.
OG and AT acknowledge support by Fermi Cycle 14 Guest Investigator program 80NSSC22K0031.
DI is supported by Future Investigators in NASA Earth and Space Science and Technology (FINESST) award No. 80NSSC21K1851. %Finesst
JJ and AT acknowledge support by NSF grants AST-2009884 %NSF neutrino factories
and NASA 80NSSC21K1746 %NASA ns accretion
grants.
AT and FF acknowledge support from 
NSF grant AST-2107839 %new short GRB grant
and NASA grant 80NSSC18K0565. %neutrino francois
AT was also supported by NSF grants
AST-1815304, %old short GRB grant
AST-1911080, %accretion grant
AST-2206471, %tidal disruptions
OAC-2031997.  %Frontera travel grant
AT was also partly supported by an NSF-BSF grant 2020747.
FF also acknowledges support from the Department of Energy, Office of Science, Office of Nuclear Physics, under contract No. DE-AC02-05CH11231 and NASA through grant 80NSSC22K0719.
RP acknowledges support by NSF award AST-2006839.
MD acknowledges support from PHY-2110287.
Support for this work was also provided by the National Aeronautics and Space Administration through Chandra Award Number TM1-22005X issued by the Chandra X-ray Center, which is operated by the Smithsonian Astrophysical Observatory for and on behalf of the National Aeronautics Space Administration under contract NAS8-03060.
This research used resources of the Oak Ridge Leadership Computing Facility, which is a DOE Office of Science User Facility supported under contract DE-AC05-00OR22725.  This research was facilitated by the Multimessenger Plasma Physics Center (MPPC), NSF grant PHY-2206607.  This research used resources of the National Energy Research Scientific Computing Center, a DOE Office of Science User Facility supported by the Office of Science of the U.S. Department of Energy under contract No. DE-AC02-05CH11231 using NERSC award NP-ERCAP0020543 (allocation m2401).
An award of computer time was provided by the ASCR Leadership Computing Challenge (ALCC), Innovative and Novel Computational Impact on Theory and Experiment (INCITE), and OLCF Director's Discretionary Allocation  programs under award PHY129. This research used resources of the National Energy Research Scientific Computing Center, a DOE Office of Science User Facility supported by the Office of Science of the U.S. Department of Energy under contract No. DE-AC02-05CH11231 using NERSC award ALCC-ERCAP0022634.
    
\end{acknowledgements}

\section*{Data Availability}

The data underlying this article will be shared upon reasonable request to the corresponding author.

\bibliography{refs}

\appendix

\section{Remapping SpEC to \textsc{h-amr}} \label{sec:remap}

In our attempt to self-consistently simulate the BH--NS merger, starting from the inspiral all the way to $ r > 10^{11}\,{\rm cm} $, we employ a novel approach in which we use the outcome of the dynamical spacetime simulation, using code SpEC, as an initial setup. We remap the set of primitive quantities, such as density, pressure and velocity, onto the grid of the GPU-accelerated GRMHD code \textsc{h-amr}, which employs a static spacetime that allows us to greatly reduce the computational cost of the simulations, and simulate the entire dynamical evolution timescale of the post-merger ejecta.

The final snapshot of the BH--NS simulation includes a set of the primitive quantities (density, pressure, and plasma four-velocities), the covariant metric components, denoted as $g_{\mu\nu}^{\rm(SpEC)}$, and the inertial coordinates of the grid of the simulation $x_{\rm SpEC}$. The grid of our simulation employs a Kerr metric in horizon-penetrating modified Kerr-Schild coordinates, $(\log r, \theta, \varphi)$. The inertial coordinates of the dynamical spacetime simulation asymptote to Cartesian Kerr-Schild coordinates far away from the BH. Therefore, we first use a Cartesian to spherical polar coordinate transformation. However, near the BH, we use a special procedure to find the coordinate transformation, since the metric at this region differs considerably from the Kerr-Cartesian metric.
We use the symmetric nature of the covariant metric to eigendecompose the input and output metric matrices, $Q$ and $\hat{Q}$, respectively,
\begin{equation}
    g^{\rm (SpEC)} = \hat{Q}\hat{\Lambda}\hat{Q}^T, \quad g^{\rm (H-AMR)} = Q\Lambda Q^T\,,
    \label{eqn:symm_matrix}
\end{equation}
where the columns of $Q$ and $\hat{Q}$ are normalized eigenvectors, and $\Lambda$ and $\hat{\Lambda}$ are diagonal matrices with eigenvalues ordered from the smallest to the largest.

The coordinate transformation matrix from SpEC to \textsc{h-amr} coordinates, $J^{\hat{a}}_{a} = \frac{\partial x^{\hat{a}}_{\rm SpEC}}{\partial x^a_{\rm H-AMR}}$, connects the metrics as follows:
\begin{equation}
g^{\rm(SpEC)}_{\hat{\mu}\hat{\nu}} \frac{\partial x^{\hat{\mu}}_{\rm SpEC}}{\partial x^\mu_{\rm H-AMR}} \frac{\partial x^{\hat{\nu}}_{\rm SpEC}}{\partial x^\nu_{\rm H-AMR}} = g^{\rm (H-AMR)}_{\mu\nu}\,,
    \label{eqn:gcov_spec2hamr}
\end{equation}
or, in a matrix form,
\begin{equation}
    J^T g^{\rm(SpEC)} J = g^{\rm (H-AMR)}\,.
    \label{eqn:jacobian}
\end{equation}
Since the metric matrices have the metric signature that is preserved after the coordinate transformation (i.e. there is always one negative and three positive eigenvalues), we further decompose the eigenvalue matrix as the product of the matrices $\Lambda = RDR$, where $R_{ii} = \sqrt{|\Lambda_{ii}|}$ and $D=\rm diag (-1,1,1,1)$. 
Together with Equations~\eqref{eqn:symm_matrix},\eqref{eqn:jacobian}, we obtain the matrix equation (where $D = \hat{D}$)
\begin{equation}
    J^T \hat{Q}\hat{R}\hat{D}\hat{R}\hat{Q}^T J = Q R D R Q^T\,.
\end{equation}
The coordinate transformation matrix is thus
\begin{equation}
    J = (\hat{R}\hat{Q}^T)^{-1} R Q^T = \hat{Q} \hat{R}^{-1} R Q^T\,.
\end{equation}

By computing the eigenvalues and eigenvectors of the metric matrices, we compute the coordinate transformation at each point, which preserves the inner product of the four-velocity by default. We use the numerical implementation of the eigendecomposition provided by the NumPy package in Python \citep[specifically, the eigh() function;][]{Harris2020}. The major caveat here is that the eigendecompositions of symmetric matrices are unique up to (a) the signs of the normalized eigenvectors and (b) eigenvector-eigenvalue pair column permutations. Therefore, we develop a numerical procedure that (a) guarantees that the eigenvectors are continuous in sign and (b) minimizes the matrix difference between the SpEC and \textsc{h-amr} metric matrix eigendecompositions using the Magyar algorithm. Additionally, we rescale the cylindrical radius in the final coordinate system $r_{\rm cyl} = \int \sqrt{g_{\varphi\varphi}}d\varphi/2\pi$, to match the cylindrical radius computed in the SpEC coordinates. We map the values of density and pressure and use the coordinate transformation matrix at each point on the grid to remap the velocities,
\begin{equation}
    u^{\mu}_{\rm H-AMR} = J^T u^{\hat{\mu}}_{\rm SpEC}\,.    \label{eqn:velocityremap_spec2hamr}
\end{equation}

\begin{figure}
    \centering    	\includegraphics[scale=0.15]{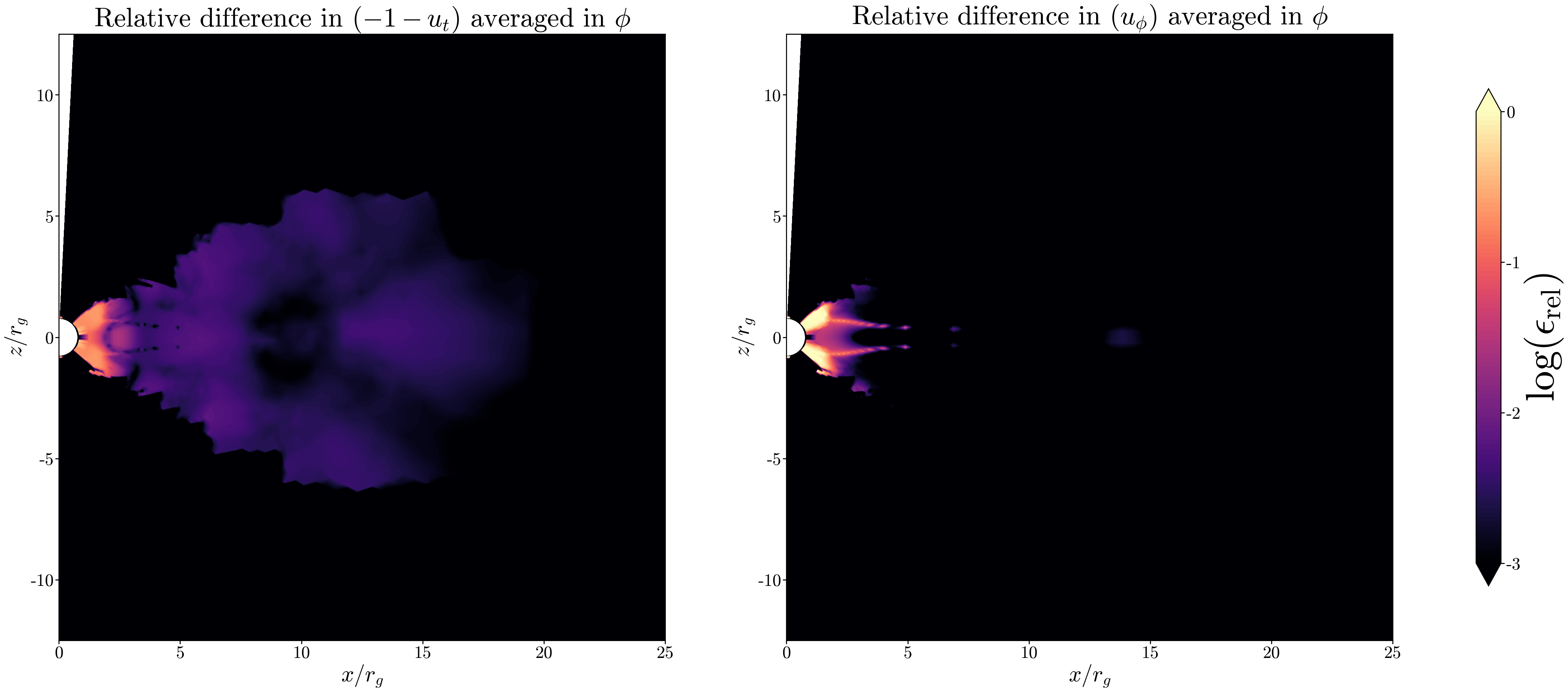}
     \caption{Relative difference in the $\varphi$-averaged specific internal energy (left) and specific angular momentum (right), before and after the remapping procedure. The largest differences, which do not exceed $\sim 10 \%$, are seen in the region surrounding the BH, where the SpEC metric differs the most from the time-independent Kerr metric, and the difference is even smaller at larger radii.
     %{\FF Is this a log scale? The colorbar needs a legend.}
     }
    \label{fig:remap_quantities}
    \end{figure}

We verify the handoff procedure by comparing the profiles of $\varphi$-averaged specific binding energy ($-1-u_t$) and angular momenta ($u_\varphi$), as seen in Figure~\ref{fig:remap_quantities}, which differ by $\lesssim 10\%$ at in the regions near the BH ($r\lesssim 3\,r_g$).

\begin{figure}
    \centering    	
    \includegraphics[scale=0.24]{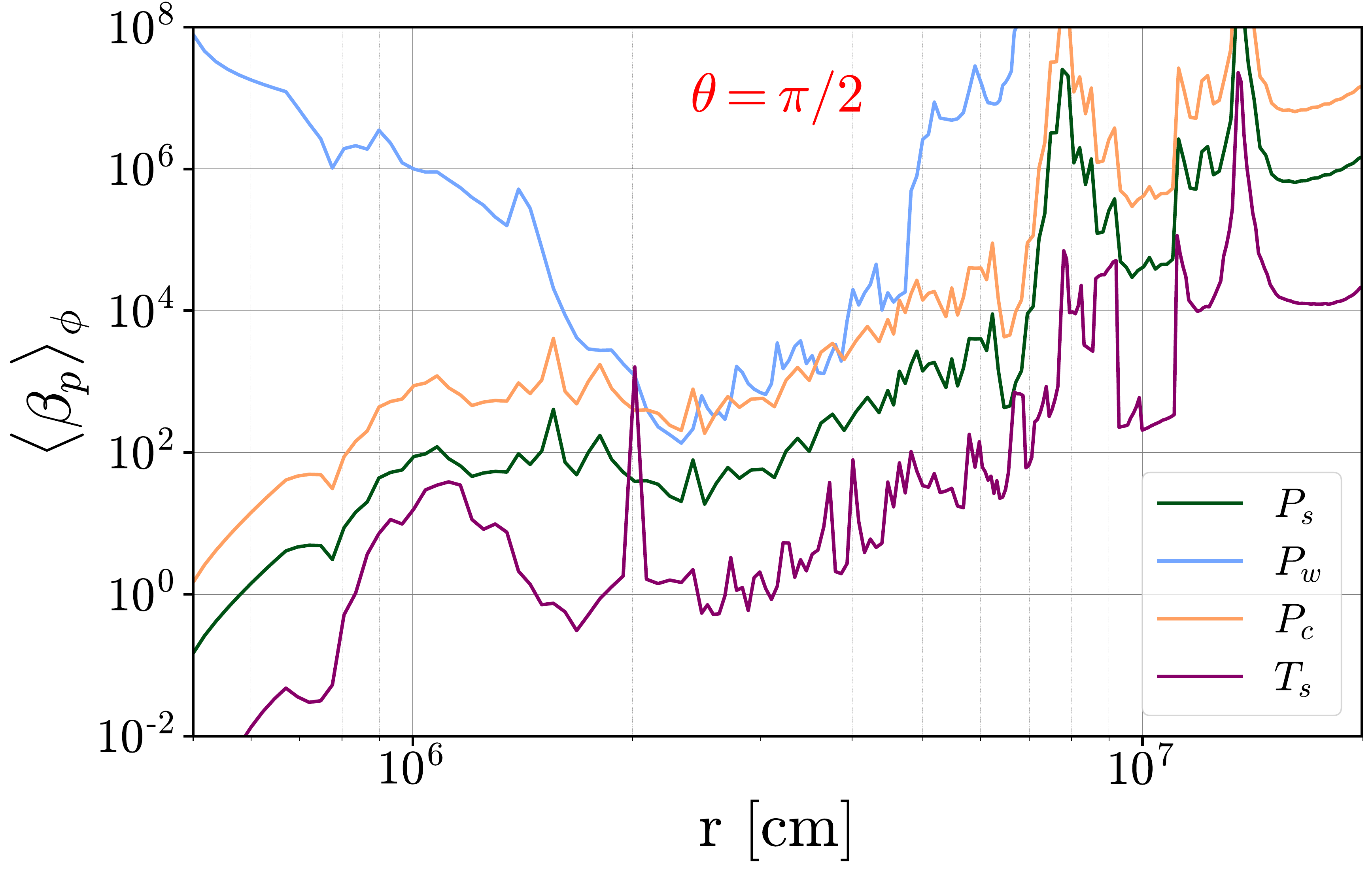}
     \caption{Azimuthal angle ($\varphi$) averaged $ \beta_p $ on the equator at the start of the simulation for each of the models, weighted by density.}
    \label{fig:betap_ini}
    \end{figure}

At the time of the remapping, we introduce various magnetic field profiles for gas with $ \rho > 5\times 10^{-4}\rho_{\rm max} $ (see Tab.~\ref{tab:models}), where $ \rho_{\rm max} $ is the maximum comoving mass density in the grid at the time of remapping.
Figure~\ref{fig:betap_ini} depicts the radial profile of the initial mass density weighted average $ \beta_p $ on the equatorial plane.

\section{Convergence test} \label{sec:convergence}

\begin{figure}
    \centering    	
    \includegraphics[scale=0.28]{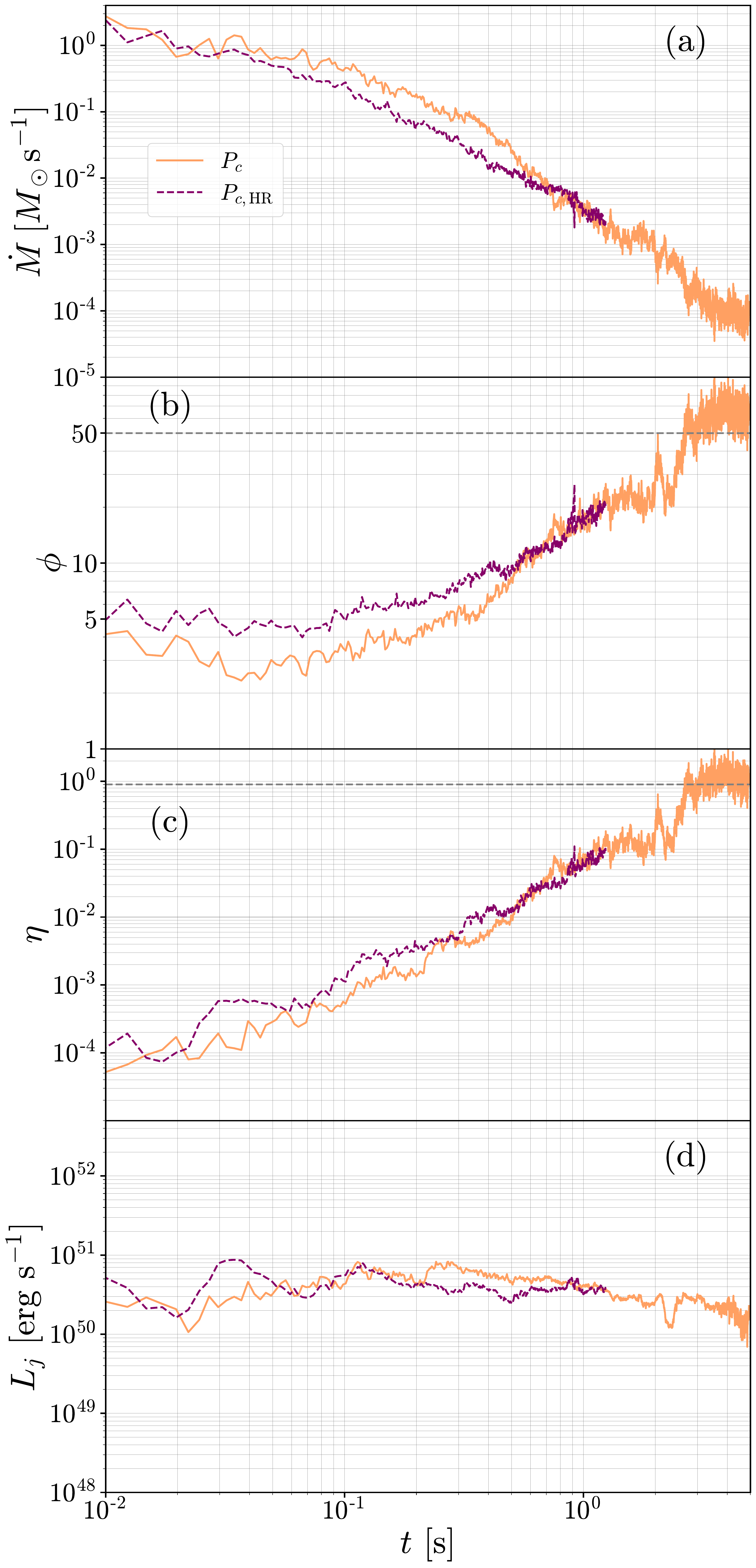}
     \caption{Same as Fig.~\ref{fig:launching}, but for model configuration $ P_c $ with different resolutions: original ($ P_c $) and high resolution ($ P_{c,\,{\rm HR}} $) where the resolution in the disk is doubled in all three dimensions. Negligible differences are seen at all times in all quantities, indicating that the simulations are converged.
     }
    \label{fig:convergence}
    \end{figure}

We verify that our simulations converge by performing one high resolution simulation of configuration $ P_c $. In that simulation we double the resolution in the disk in all dimensions, and compare it with the original resolution. Figure~\ref{fig:convergence} depicts the same quantities as in Fig.~\ref{fig:launching}, but for the two resolutions. All quantities are compatible between the two resolutions at all times. This confirms that the disk evolution and jet launching, including the MRI, are well resolved in our original resolution.

\section{Hydrodynamic disk evolution} \label{sec:hydro_disk}
\begin{figure}
    \centering    	\includegraphics[scale=0.35]{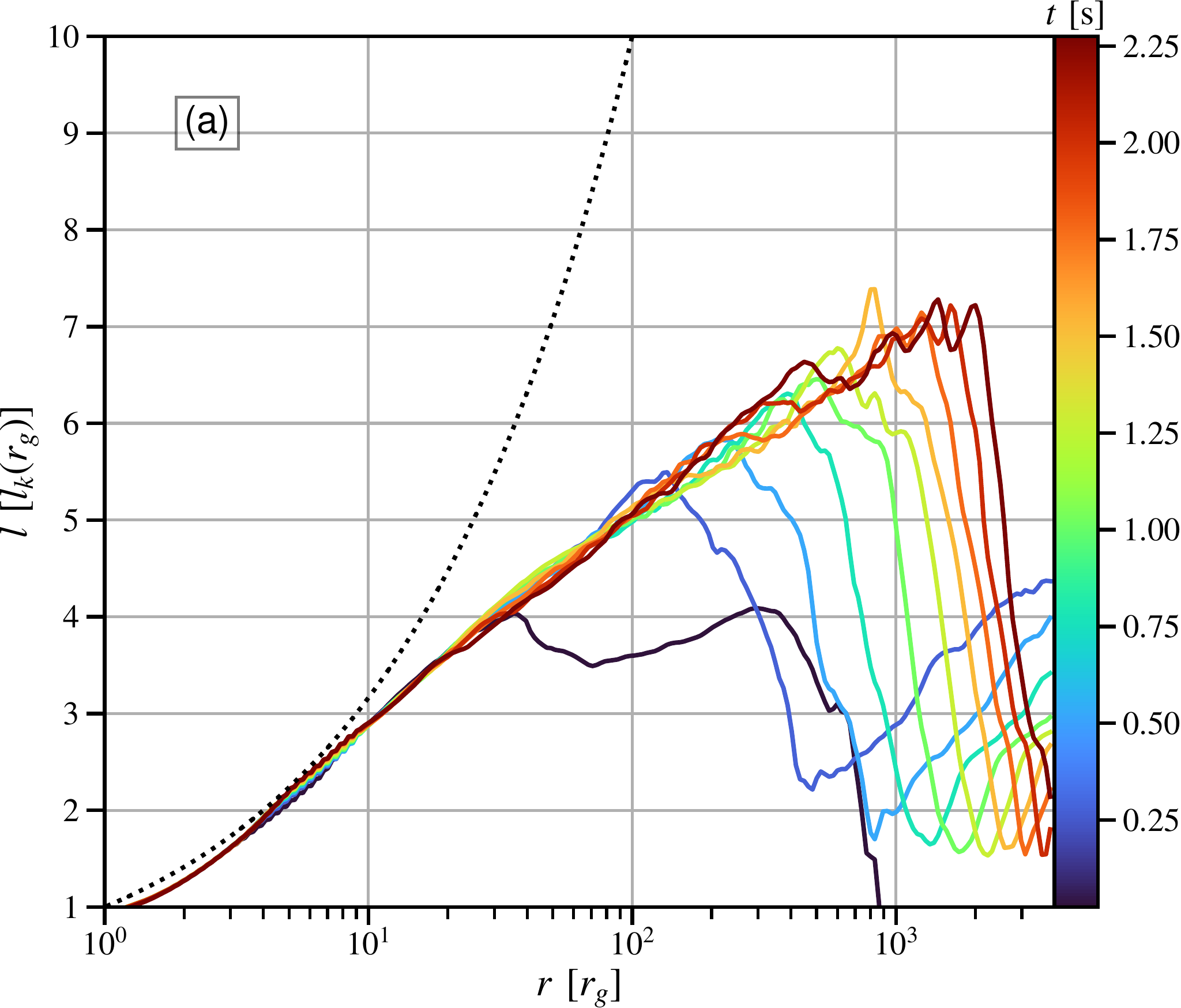}
    \includegraphics[scale=0.35]{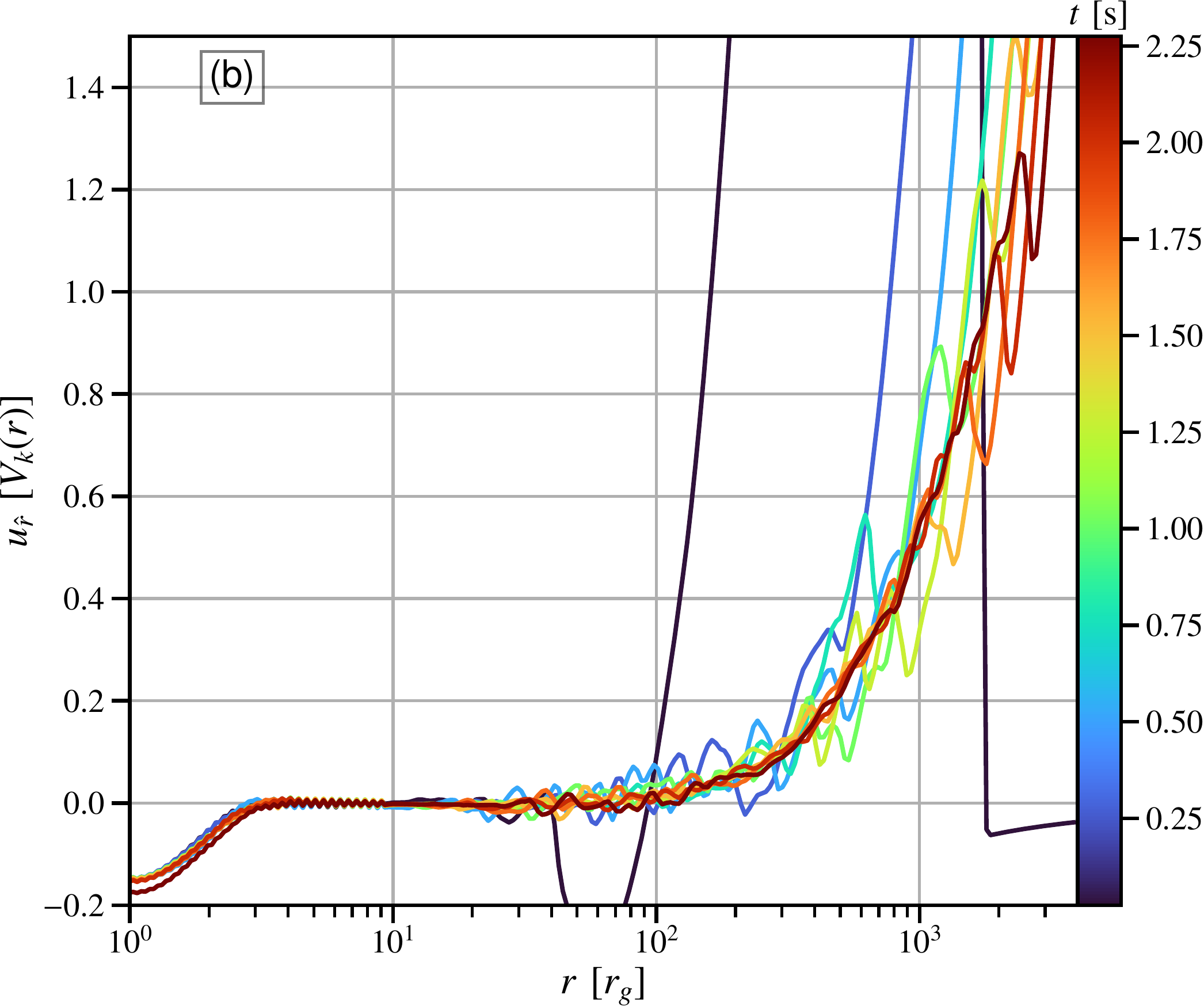}
     \caption{{\bf Panel (a)}: Radial profile of the density averaged angular momentum, $l$, normalized to the non-relativistic expression of the Keplerian angular momentum at $r_g$ at different times. The Keplerian angular momentum profile, $l_k(r)\propto r^{1/2}$ (dotted black line) illustrates that the angular momentum profile is always sub-Keplerian. {\bf Panel (b)}: Radial profile of the density averaged physical radial velocity, $u_{\hat{r}} = \sqrt{g_{rr}}u^r$, normalized to the local Keplerian velocity, $V_K(r)\propto r^{-1/2}$, at different times. The radial flux of angular momentum drives accretion in the inner regions and radial outflow in the outer regions.
     }
    \label{fig:velocity}
    \end{figure}

\begin{figure}
    \centering    	\includegraphics[scale=0.14]{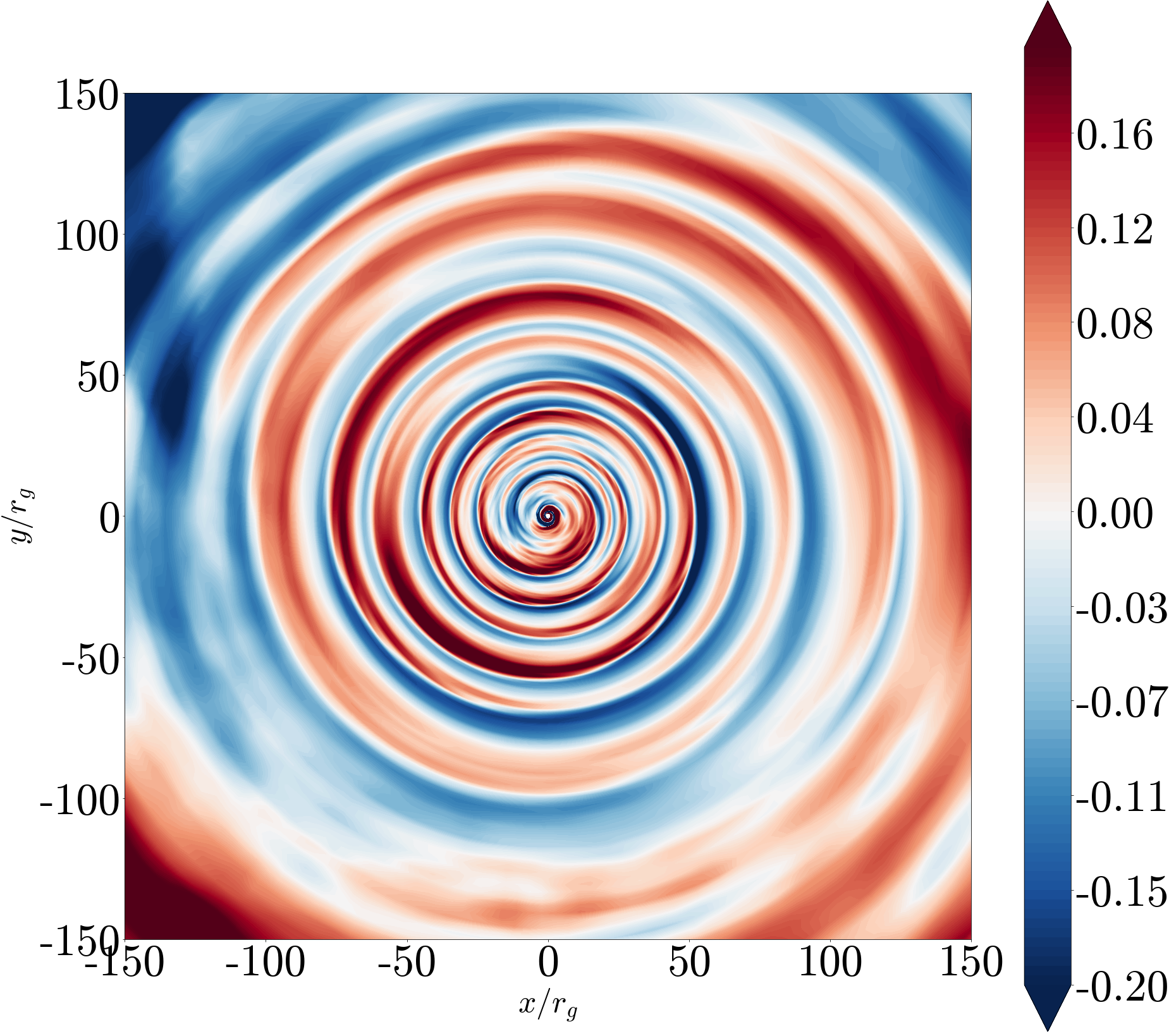}
     \caption{Relative deviations from azimuthally-averaged surface density profile, ${\Sigma}/{\overline{\Sigma}} - 1$, taken $ 0.5\,{\rm s} $ after the merger. Here, $\Sigma(r,\varphi)$ is the surface density, and $\overline{\Sigma}(r)$ is the azimuthally averaged surface density.}
    \label{fig:spiral}
    \end{figure}

Figure~\ref{fig:velocity}(a) depicts the density averaged radial angular momentum profile at different times. All profiles deviate considerably from Keplerian (black dotted line) at $ r \gtrsim 20\,r_g $. This implies that the disk is pressure supported at the outer radii. The profile at the earliest time (dark blue) is relatively flat, similar to a thick torus, and is thus a fertile ground for the Papaloizou–Pringle instability \citep[PPI,][]{papaloizou_dynamical_1984}. Fig.~\ref{fig:velocity}(a) demonstrates a slow readjustment of the angular momentum profile, as angular momentum is progressively transported towards the outer radii. This angular momentum transport drives accretion in the inner radii. Figure~\ref{fig:spiral} delineates deviations of the surface density from an axisymmetric profile. It portrays a spiral structure in the disk, which is the underlying mechanism for transport of angular momentum, matter and enthalpy (in the form of dissipating shocks) to the outer radii. 

Fig.~\ref{fig:velocity}(b) shows the density averaged radial physical velocity, $u_{\hat{r}} = \sqrt{g_{rr}}u^r$, normalized to the local Keplerian velocity. The positive radial flux of angular momentum leads to accretion in the inner radii and outward radial motions. The radial oscillations in the radial velocity, driven by the spiral shocks, are damped as the disk redistributes its angular momentum, and the PPI stabilizes itself on sufficiently long timescales \citep{bugli_papaloizoupringle_2018}. The radial velocity is comparable to the Keplerian velocity at $r>10^{3}\,r_g $, effectively reaching escape velocity, which explains the outflow measured in \S\ref{sec:ejecta}.

In the absence of neutrino cooling in the system, the torus can only cool through vertical transport. Thus, the radial transport also drives a vertical outflow. Neutrino cooling would efficiently inhibit the outward radial transport of enthalpy by vertically disposing the thermal energy. Finally, MRI turbulence is known to stabilize the PPI, and is also expected to dominate the angular momentum transport \citep{bugli_papaloizoupringle_2018}. Thus, accretion through PPI is also not physically motivated in BH--NS mergers.

\end{document}